\documentclass[journal]{IEEEtran}

\usepackage{graphicx}
\usepackage{mathtools}
\usepackage{cuted}
\usepackage{eqnarray, amsmath}
\allowdisplaybreaks
\usepackage{amssymb}
\usepackage{multicol,lipsum}
\usepackage{cite}
\usepackage[colorlinks=Flase]{hyperref}
\usepackage{bbm}
\usepackage{algorithm} 
\usepackage{algpseudocode}
\usepackage{caption}
\usepackage{steinmetz}
\usepackage{subcaption}
\usepackage{amsthm}
\usepackage[english]{babel}

\newtheorem{lemma}{Lemma}

\newcounter{relctr} 
\everydisplay\expandafter{\the\everydisplay\setcounter{relctr}{0}} 
\DeclareMathOperator*{\minimize}{minimize} 
\DeclareMathOperator*{\maximize}{maximize} 
\usepackage[figurename=Fig.]{caption}

\newtheoremstyle{remarkstyle}%
  {}
  {}
  {\itshape}
  {}
  {\itshape}
  {.}
  {.5em}
  {}

\theoremstyle{remarkstyle}
\newtheorem{remark}{Remark}

\AtBeginDocument{} 
\newcommand{\norm}[1]{\left\lVert#1\right\rVert}

\ifCLASSINFOpdf
\else
\fi

\begin{document}

\title{On the Deployment of Multiple Radio Stripes for Large-Scale Near-Field RF Wireless Power Transfer}

\author{Amirhossein~Azarbahram,~\IEEEmembership{Graduate~Student~Member,~IEEE,}
        Onel~L.~A.~López,~\IEEEmembership{Senior~Member,~IEEE}
        Petar~Popovski,~\IEEEmembership{Fellow,~IEEE,}
        and~Matti~Latva-Aho,~\IEEEmembership{Fellow,~IEEE}
\thanks{A. Azarbahram, O. L\'opez and M. Latva-Aho are with Centre for Wireless Communications (CWC), University of Oulu, Finland, (e-mail: \{amirhossein.azarbahram, onel.alcarazlopez, matti.latva-aho\}@oulu.fi). P.~Popovski is with the Department of Electronic Systems, Aalborg University, 9100 Aalborg, Denmark (e-mail: petarp@es.aau.dk).}%
\thanks{This work is partially supported in Finland by the Research Council of Finland (Grants 348515, 362782, and 369116 (6G Flagship)); by the European Commission through the Horizon Europe/JU SNS project AMBIENT-6G (Grant 101192113); and in Denmark by the Villum Investigator Grant “WATER” from the Velux Foundation (Grant 37793).}
}


\maketitle

\begin{abstract}

This paper investigates the deployment of radio stripe systems for indoor radio-frequency (RF) wireless power transfer (WPT) in line-of-sight near-field scenarios. The focus is on environments where energy demand is concentrated in specific areas, referred to as 'hotspots', spatial zones with higher user density or consistent energy requirements. We formulate a joint clustering and radio stripe deployment problem that aims to maximize the minimum received power across all hotspots. To address the complexity, we decouple the problem into two stages: i) clustering for assigning radio stripes to hotspots based on their spatial positions and near-field propagation characteristics, and ii) antenna element placement optimization. In particular, we propose four radio stripe deployment algorithms. Two are based on general successive convex approximation (SCA) and signomial programming (SGP) methods. The other two are shape-constrained solutions where antenna elements are arranged along either straight lines or regular polygons, enabling simpler deployment. Numerical results show that the proposed clustering method converges effectively, with Chebyshev initialization significantly outperforming random initialization. The optimized deployments consistently outperform baseline benchmarks across a wide range of frequencies and radio stripe lengths, while the polygon-shaped deployment achieves better performance compared to other approaches. Meanwhile, the line-shaped deployment demonstrates an advantage under high boresight gain settings, benefiting from increased spatial diversity and broader angular coverage.

\end{abstract}


\IEEEpeerreviewmaketitle

\section{Introduction}

\IEEEPARstart{R}{adio} frequency (RF) wireless power transfer (WPT) is emerging as a key enabler of battery-free operation and reduced maintenance in future wireless networks. Unlike other WPT technologies, RF-WPT can simultaneously charge multiple devices through controlled radiation and integrate seamlessly with existing wireless infrastructure \cite{lópez2023highpower, ZEDHEXA}. Interestingly, RF-WPT, referred to as WPT throughout this work, does not require a dedicated energy source; instead, RF power can be transmitted over the same infrastructure used for wireless communications. However, the main drawback of WPT is its low end-to-end efficiency, primarily due to severe wireless channel losses. Since these losses increase rapidly with distance, long-range WPT remains highly challenging. This has motivated advanced techniques such as energy beamforming and distributed antenna architectures to improve power delivery over extended ranges \cite{intro3}.

Beamforming can mitigate channel losses by focusing energy on the devices \cite{intro3}. However, conventional multi-antenna systems with co-located elements still suffer from significant path loss, since the elements experience nearly identical propagation paths. Consequently, blockages or large user distances affect all antennas similarly, and only very large apertures could overcome this, often impractical in practice. To address non-line-of-sight limitations, reflective reconfigurable intelligent surfaces (RIS) with passive elements can be deployed \cite{IRS-basis}. However, the performance of RIS can be limited by the attenuation between the transmitter and the reflecting elements. As a more active alternative, distributed antenna systems (DAS) can be deployed across the area to reduce blind spots and shorten charging distances, thereby improving WPT efficiency \cite{DAS_ref_heath}. Another key advantage of employing DAS is their ability to provide a large aperture diameter, which is often constrained in conventional co-located antenna arrays. As a result, DAS are more capable of supporting near-field transmission, where the wavefronts arriving at a receiver exhibit a spherical nature. Operating in the near-field enables focusing RF power toward specific spatial locations, rather than forming the beams in certain directions, which can lead to more precise and efficient energy delivery in WPT systems \cite{near-field}. Recently, hybrid solutions have been proposed that combine distributed active antennas with passive reflecting surfaces to further improve the coverage \cite{dist_RIS}. Another possible solution for distributing the elements is utilizing unmanned aerial vehicle (UAV)-assisted systems. UAVs can provide both direct and short-distance energy transmission paths; thus, a UAV-assisted system with a proper trajectory design may significantly reduce the average channel loss in the system \cite{UAVzhang2018}. However, the energy required for UAV movement and hovering can considerably reduce the end-to-end power transfer efficiency, especially over longer durations or in dynamic environments. Moreover, fairness remains a key challenge in such systems, as relocating a UAV (or focusing energy) toward one user can significantly degrade power transfer to others, possibly leading to coverage gaps \cite{uav_survey}. Therefore, it is essential to optimize the placement of transmit antennas based on the spatial characteristics of the deployment area and user distribution. 

Interestingly, radio stripe systems offer a cost-effective DAS solution in large areas. In this architecture, antenna elements and their associated processing units are embedded along a cable, enabling streamlined deployment with minimal infrastructure requirements. A sufficiently long radio stripe can accommodate a large number of antenna elements dispersed throughout the environment, thereby reducing the average distance to users and mitigating blockage issues \cite{intro_radiostripe}.

\subsection{Prior Works}

Several studies in the literature address the deployment of DAS for WPT to enhance system efficiency and coverage. For example, the authors in \cite{distributedWPT1} propose an analytical framework to determine the optimal radius of a circular layout where power beacons (PBs) are uniformly distributed, to maximize WPT efficiency. In \cite{intro3}, the impact of the number of PBs on the minimum received energy is analyzed, and K-means clustering is employed to optimize PB placement. Furthermore, \cite{Osmel_Deployment} explores multiple deployment strategies to position PBs such that the average energy harvested at the worst-case location is maximized. In \cite{deploy_WPCN}, a non-uniform deployment strategy for PBs is proposed, where placement density is adapted based on user locations and energy demand. This approach demonstrates that adaptive, spatially aware distributions lead to higher WPT efficiency than uniform designs. The work in \cite{WPT_dist_survey} presents a comprehensive design and prototype of a distributed antenna WPT system, demonstrating experimentally how spatial diversity and antenna selection can be leveraged in practical deployments to improve the coverage. The study in \cite{twoantennaWPT} investigates the optimal placement of two transmit antennas in near-field WPT and demonstrates that even simple antenna configurations can significantly affect power delivery depending on the user’s spatial distribution. Moreover, the results emphasize the critical role of antenna deployment in near-field scenarios, showing that while co-located antennas are preferable in the far-field, distributed antenna placements are more effective in the near-field. Furthermore, \cite{WPT_PB_indoor} studies antenna deployment and power allocation in indoor environments, showing that strategic placement within the radiating near-field region significantly enhances efficiency under practical constraints. Finally, \cite{physicallylargeapperture} experimentally demonstrates that utilizing physically large apertures, such as radio stripes, can significantly enhance the efficiency and safety of WPT.

While some prior works \cite{distributedWPT1, WPT_dist_survey, WPT_PB_indoor} have explored DAS for WPT, and others \cite{OnelRadioStripes} proposed beamforming techniques for radio stripe-based energy delivery, no work has yet addressed the deployment optimization of radio stripes under near-field WPT conditions. Future WPT applications are expected to take place predominantly in indoor environments, where power demand is concentrated around predefined hotspots. In such scenarios, radio stripes offer a scalable and cost-effective means to distribute transmit antennas, and the efficiency of power delivery highly depends on how the antenna elements are deployed. Unlike far-field scenarios, near-field WPT enables power focusing at specific spatial locations, making antenna deployment even more critical \cite{twoantennaWPT}. Hence, optimizing the radio stripe deployment becomes fundamental to unlock the full potential of near-field safe WPT.

\subsection{Contributions}

\begin{figure}
    \centering
    \includegraphics[width=\columnwidth]{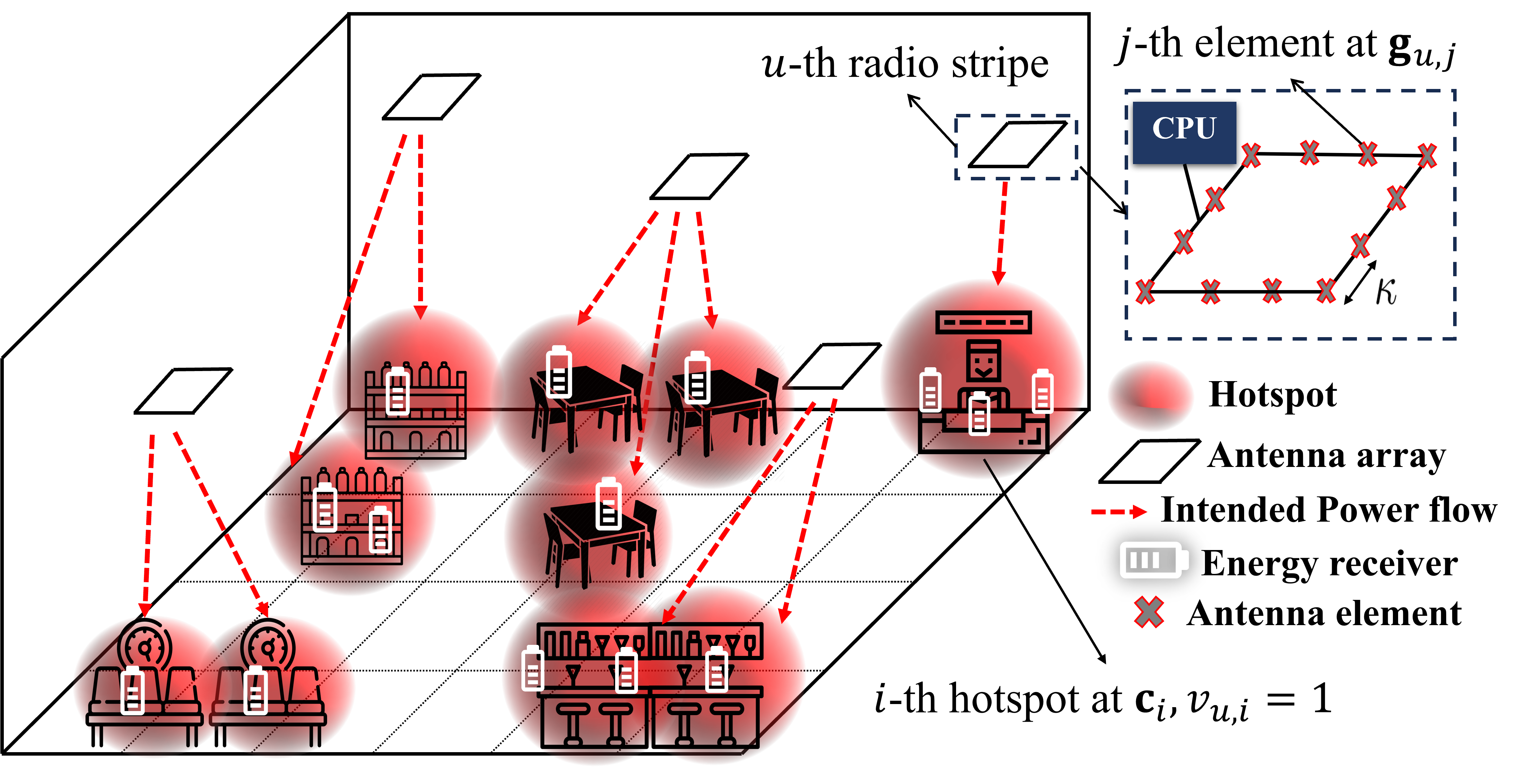}
    \caption{Distributed radio stripe system model, exemplified with a shopping mall scenario. Moreover, each radio stripe system is connected to one central processing unit (CPU). }
    \label{fig:system_model}
\end{figure}
 
Herein, we consider, for example, a shopping mall as in Fig.~\ref{fig:system_model}, where devices like wearables are typically placed near tables or service counters, forming localized hotspots. Our main contributions are:\footnote{This work is an extended version of the conference paper \cite{azarbahram2023radio}, which presents an initial study on radio stripe deployment for indoor WPT.}

\par \textit{\textbf{First}}, we formulate a joint clustering, beamforming design, and radio stripe deployment problem for a near-field WPT system, aiming to maximize the minimum received power across multiple hotspot locations. In this setup, each radio stripe acts as a cluster head, serving a group of hotspots. In practice, the clustering and deployment are performed offline during the planning phase, while beamforming can be adapted at runtime based on local channel state information (CSI) or by relying solely on the spatial distribution of devices.

\par \textit{\textbf{Second}}, we propose a two-stage decoupled solution approach, which solves the clustering problem, followed by joint beamforming and antenna element location optimization within each radio stripe. For clustering, we introduce a loss metric based on the gain of the near-field line-of-sight (LoS) channel and formulate an alternating optimization algorithm to jointly determine the user-to-cluster assignments and the optimal positions of the cluster heads, which match the radio stripes centers. In the second stage, we adopt maximum ratio transmission (MRT)-based precoders and eliminate the impact of non-dedicated beams to reduce the complexity of beamforming optimization. We then propose four deployment strategies: two high-complexity methods based on successive convex approximation (SCA) and signomial programming (SGP), and two low-complexity SGP-based solutions that restrict the radio stripe layout to predefined line and polygon shapes. Furthermore, we develop a heuristic mapping procedure to convert the continuous optimization outputs of the SCA and SGP methods into feasible layouts that comply with the practical inter-element distance constraints of radio stripes.

\par \textit{\textbf{Third}}, we show numerically that:\\
i) The proposed clustering algorithm converges reliably and consistently outperforms the baseline K-Chebyshev method, which assigns users based on Chebyshev distances.\\
ii) All radio stripe deployment schemes converge, with the SCA-based approach requiring more iterations due to its alternating structure.\\
iii) The proposed deployments outperform benchmarks across different frequencies and radio stripe lengths for both MRT- and SDP-based precoders, with polygon-shaped layouts achieving the best performance.\\
iv) At high boresight gain values, the line-shaped deployment benefits from greater angular diversity and eventually outperforms the other strategies.

The remainder of the paper is organized as follows. Section~\ref{sys and prob} presents the system model and formulates the joint optimization problem. Section~\ref{sec:Optimization} describes the proposed optimization framework. Low-complexity radio stripe deployment methods are detailed in Section~\ref{sec:low_comp}, while Section~\ref{sec:converge} analyzes the convergence behavior and computational complexity of the proposed algorithms. Numerical results are provided in Section~\ref{result}, and the paper is concluded in Section~\ref{conclusion}.

\textbf{Notations}: Bold lowercase and uppercase letters represent vectors and matrices, respectively, while the $l$th element in a vector is denoted by $\mathbf{a}[l]$. The notation $\norm{\cdot }$ denotes the $l_2$ norm of a vector, while $(\cdot)^T$ and $(\cdot)^H$ indicate the transpose and transpose conjugate of a matrix or vector, respectively. 

\section{System Model \& Problem Formulation}\label{sys and prob}

We consider a multi-antenna WPT system comprising $U$ radio stripe networks that are deployed to supply energy to $K$ designated hotspots, which are spatial regions where energy demand is concentrated. These hotspots are assumed to be known in advance and remain static during the deployment planning phase, while the devices are located in the proximity of the hotspot center. This assumption is not restrictive, since radio planning in cellular networks is performed under similar conditions, where traffic demand hotspots are forecasted and used as the basis for transmitter deployment. Each radio stripe network is assigned to a distinct subset of hotspots, referred to as a cluster, which it serves independently from other stripes. To represent the association between hotspots and radio stripes, we define a binary variable $v_{u,i} \in \{0, 1\}$, where $v_{u,i} = 1$ indicates that the $u$-th radio stripe network serves the $i$-th hotspot, and $v_{u,i} = 0$ otherwise. Moreover, each radio stripe network consists of $N$ elements with an inter-element distance of $\kappa$, which are located on the ceiling with height $h_c$. Furthermore, $\mathbf{g}_{u, j}$ is the Cartesian coordinate of the $j$-th antenna element in the $u$-th radio stripe network, and $\mathbf{c}_{i}$ denotes the location of the center of the $i$-th hotspot. Notably, the antenna elements of each radio stripe are physically embedded along a common cable, connected sequentially. The system model depicted in Figure~\ref{fig:system_model} depicts a shopping mall configuration, featuring designated hotspot areas where the devices are expected to be positioned.

\begin{remark}
We assume that radio stripes operate without information exchange, with each transmitter serving only its associated users. This corresponds to a non-coordinated multi-point transmission setup, which simplifies system design and reduces signaling overhead, but may incur some performance loss compared to fully coordinated schemes.
\end{remark}
\subsection{Channel Model \& Received Power}

Given our emphasis on WPT within indoor environments, we consider a near-field LoS wireless channel model \cite{near-field}. Let $D$ be the antenna array diameter, then user $i$ at a distance $r_i$ from the transmitter lies in the radiative near-field region if
\begin{equation}\label{eq:nearfield}
    \sqrt[3]{\frac{D^4}{8\lambda}} = r_{fs} < r_i < r_{fr} = \frac{2D^2}{\lambda},
\end{equation}
where $r_{fs}$ and $r_{fr}$ are the Fresnel and Fraunhofer distance, respectively. Thus, it is evident that the near-field region can be expanded by increasing both the system frequency and the size of the antenna array. For arbitrary radio stripe geometries, we define the diameter of the $u$-th radio stripe ($D_u$) as the maximum Euclidean distance between any two antenna elements within the $u$-th radio stripe, i.e., 
\begin{equation}\label{eq:Du}
    D_u = \max_{j,k} \norm{\mathbf{g}_{u,j} - \mathbf{g}_{u,k}}.
\end{equation}

The channel coefficient between the $j$-th element of the $u$-th radio stripe and hotspot $i$  is given by \cite{near-field}
\begin{equation}\label{eq:channelcoef}
    h_{u, j, i} = A_{u, j, i} e^{\frac{-j2\pi}{\lambda} \norm{\mathbf{g}_{u, j} - \mathbf{c}_{i}}},
\end{equation}
where the term $2\pi ||\mathbf{g}_{u, j} - \mathbf{c}_{i}||/\lambda$ represents the phase shift that is introduced due to the propagation distance and $\lambda$ is the wavelength. Moreover, $A_{u, j, i}$ is the corresponding channel gain, which is given by
\begin{equation}
    A_{u, j, i} = \sqrt{F(\theta_{u, j, i})}\frac{\lambda}{4\pi ||\mathbf{g}_{u, j} - \mathbf{c}_{i}||}.
\end{equation}
Herein, $F(\theta_{u, j, i})$ is the antenna radiation profile given by \cite{anetnna_radiation}
\begin{equation}
        F(\theta_{u, j, i}) = \begin{cases}
        2(b + 1){\cos^b {\theta_{u, j, i}}}, & \theta_{u, j, i} \in [0,\pi/2], 
        \\
        0, & \text{otherwise},
        \end{cases}
\end{equation}
where $b$ is the the boresight gain and $\theta_{u, j, i}$ is the elevation angle between the $j$-th element and the $i$-th hotspot in the $u$-th cluster. Since the antenna elements are located at the ceiling, we can write 
\begin{equation}\label{eq:cos}
    \cos{\theta_{u, j,i}} = \frac{h_c - c_{i, 3}}{\norm{\mathbf{g}_{u, j} - \mathbf{c}_{i}}},\quad \theta_{u, j, i} \in [0, \pi/2].
\end{equation}
Additionally, $\mathbf{h}_{u, i} = [
h_{u, 1,i} ,h_{u, 2,i}, \ldots, h_{u, N,i}]^T \in \mathbb{C}^{N \times 1}$ is defined as the channel coefficients vector between hotspot $i$ and the antenna elements of the $u$-th radio stripe network. 

\begin{remark}
    The adopted near-field channel model naturally includes the far-field as a special case. In the far-field regime, the channel coefficient reduces to $A_{u,i} e^{-j\psi_{u,j,i}}$, where $A_{u,i}$ depends only on the distance between hotspot $i$ and transmitter $u$, while $\psi_{u,j,i}$ is determined by the angular direction of the hotspot and the array geometry of the $u$-th radio stripe. This highlights the generality of the LoS near-field model, which seamlessly adapts to both near- and far-field conditions.
\end{remark}

We consider $M_u\leq N$ independent and normalized energy symbols to transmit using the $u$-th radio stripe. In this context, $\mathbf{w}_{u, m} \in \mathbb{C}^{N\times 1}$ is the digital precoder corresponding to the $m$-th energy symbol in the $u$-th radio stripe, while the power of the RF signal received at the center of the $i$-th hotspot and averaged out over the signal waveform is given by
\begin{equation}\label{eq:rxpower}
    P^{rx}_{i} = \sum_{k = 1}^K \sum_{m =1}^{M_u} \norm{\mathbf{h}_{u, i}^H \mathbf{w}_{u,m}}^2.
\end{equation}

\subsection{Problem Formulation}

The objective is to optimize the physical deployment of radio stripe elements in a way that ensures fair energy delivery across designated hotspots. Specifically, we aim to maximize the minimum received RF power across all hotspots. For this, hotspot clustering and digital beamforming are included in the formulation, as they directly influence deployment decisions. Specifically, beamforming is included in the formulation to influence the spatial layout rather than to model real-time operation. This is justified by the fact that, in practice, once the radio stripes are deployed, beamforming can be dynamically adapted to serve users within each hotspot using local CSI or spatial awareness. Hence, the joint design of clustering, deployment, and beamforming serves as an offline tool to ensure that the physical layout supports fair and effective energy delivery. The resulting optimization problem is formulated as
\begin{subequations}\label{prob_basic}
\begin{align}
\label{prob_basic_a}  \quad \maximize_{{\mathbf{g}_{u, j} \mathbf{w}_{u,m}. v_{u,i}}} & \min_{i}\ {P^{rx}_{i}}/\eta_{i} \\ 
\textrm{subject to} \label{prob_basic_b} \quad & \sum_{m = 1}^{M_u} \norm{\mathbf{w}_{u,m}}^2 \leq \Tilde{P}_u, \forall k,\\
& \label{prob_basic_c} \sum_{j = 1}^{N - 1} \norm{\mathbf{g}_{u, j}\!-\! \mathbf{g}_{u,j\!+\!1}} \leq (N\!-\!1)\kappa, \forall k, \\
&  \label{prob_basic_d}\norm{\mathbf{g}_{u, j}\!-\!\mathbf{g}_{k,n}}\!\geq\!\kappa, \ \ \ \forall u,j,n\ :\ {j\neq n}, \\
& v_{u,i} \in \{0,1\}, \quad \forall u,i, \label{prob_basic_f}\\
& \sum\nolimits_u v_{u,i} = 1, \quad \forall i, \label{prob_basic_g}
\end{align}
\end{subequations}
where $\Tilde{P}_u$ denotes the transmit power budget available to the $u$-th radio stripe, while $\eta_i$ represents the user density in the $i$-th hotspot, i.e., the expected number of devices located in that region. The normalization by $\eta_i$ in the objective ensures fairness across hotspots with varying user concentrations. Herein,  constraint \eqref{prob_basic_b} ensures that each radio stripe's transmit power does not exceed its power budget. Notably, \eqref{prob_basic_c} enforces a physical limit on the total length of each radio stripe based on the number of antenna elements and their spacing. Moreover, \eqref{prob_basic_d} maintains a minimum distance $\kappa$, with $\kappa\ge \lambda/2$, between any two antenna elements, while \eqref{prob_basic_f} defines the clustering association variables as binary, and constraint \eqref{prob_basic_g} ensures that each hotspot is assigned to exactly one radio stripe.

The optimization problem in \eqref{prob_basic} is inherently non-convex and combinatorial, involving mixed-integer variables due to the binary user-to-stripe association $v_{u,i}$. Moreover, the clustering decisions directly influence both the spatial deployment of the radio stripes and the design of their beamforming strategies. Specifically, the choice of which hotspots are served by which radio stripes affects the feasible deployment geometry (through constraints \eqref{prob_basic_c}-\eqref{prob_basic_d}) and the allocation of beamforming vectors (through constraint \eqref{prob_basic_b}). As a result, the joint nature of clustering, positioning, and precoding introduces a tightly coupled optimization structure that poses significant algorithmic and computational challenges.

\section{Optimization Framework} \label{sec:Optimization}

For a given clustering configuration (i.e., fixed $v_{u,i},\ \forall u,i$) and a fixed deployment of radio stripe elements, constraints \eqref{prob_basic_c}-\eqref{prob_basic_g} can be eliminated, and the optimization problem \eqref{prob_basic} can be reformulated as an SDP \cite{onellowcomp}. Notably, the optimal precoding depends on all channel coefficients, making the joint optimization of antenna locations and beamforming highly complex and computationally intractable, especially in large-scale settings. To address this, we adopt MRT-based precoders due to their closed-form structure and reduced computational complexity. MRT aligns each beam with the conjugate of the corresponding channel vector, thereby simplifying the optimization process while still ensuring directional energy focusing. Although MRT is not globally optimal, it serves as a locally optimal solution for the beamforming subproblem \cite{OnelRadioStripes}, and thus provides a useful lower bound on the achievable performance of the full problem. This allows us to decouple the beamforming design from the deployment optimization, facilitating scalable and practical algorithmic solutions. 

Recall that beamforming is included here to guide deployment decisions. Once the antenna locations are optimized, more advanced or adaptive precoding techniques, such as CSI-aware digital beamforming, can be employed. Since this work deals with multi-antenna WPT systems with a sufficiently large $N$ in each radio stripe ($N \gg M_u, \forall u$), we assume each device is primarily served by a dedicated energy beam from its associated radio stripe, leading to $M_u = \sum_i v_{u,i}$. Specifically, $\mathbf{w}_{u, m}^* = \frac{\boldsymbol{h_{u, m}}}{\norm{\boldsymbol{h_{u, m}}}}\sqrt{P_{u, m}}$ is the $m$-th MRT-based precoder in the $u$-th radio stripe, where $P_{u, m}$ is the corresponding assigned power. Hereby, \eqref{eq:rxpower} can be reformulated as 
\begin{equation} \label{eq:rxpower2}
    {P^{rx}_{i}} = \sum_{u = 1}^U \sum_{m = 1}^{M_u}  \norm{\mathbf{h}_{u,i}^H \frac{\mathbf{h}_{u,m}}{\norm{\mathbf{h}_{u,m}}}\sqrt{P_{u, m}}}^2.
\end{equation}
Even with simple MRT, it is difficult to solve \eqref{prob_basic} because of the binary variable $v_{u,i}$ and the highly non-convex objective function. Specifically, the term $e^{\frac{-j2\pi}{\lambda} \norm{\mathbf{g}_{u, j} - \mathbf{c}_{i}}}$ is an oscillating function of the distance. Although in WPT, interference from non-dedicated beams can be beneficial due to the additive nature of RF energy, we lower bound \eqref{eq:rxpower2} as $P_{i}^{rx}> \sum_{u} v_{u,i} P_{u, i} \norm{\mathbf{h}_{u,i}}^2$. This is justified by the high spatial resolution achievable in near-field beamforming with large antenna arrays, which allows most of the energy to be concentrated on the targeted hotspot. As a result, the contribution of non-dedicated beams becomes negligible. This approximation enables a tractable formulation while preserving the dominant effects relevant for deployment optimization. Hereby, the optimization problem can be approximately rewritten as
\begin{subequations}\label{prob_MRT}
\begin{align}
\label{prob_MRT_a}  \quad \maximize_{{\mathbf{g}_{u, j}, P_{u, i} v_{u,i}}} \quad & \min_{i} \sum\nolimits_{u} v_{u,i}P_{u, i} \norm{\mathbf{h}_{u,i}}^2/\eta_{i} \\ 
\textrm{subject to} \label{prob_MRT_b} \quad & \sum\nolimits_{i} v_{u,i}P_{u, i} \leq \Tilde{P}_u, \forall u, \\
& \eqref{prob_basic_c}, \eqref{prob_basic_d}, \eqref{prob_basic_f}, \eqref{prob_basic_g}. \nonumber
\end{align}
\end{subequations}
Additionally, by utilizing \eqref{eq:channelcoef}-\eqref{eq:cos}, we can write 
\begin{align}\label{eq:gain_simp}
    \norm{\mathbf{h}_{u,i}}^2 &= \sum_{j = 1}^{N} \bigl(\sqrt{2(b + 1){\cos^b {\theta_{u, j,i}}}}\frac{\lambda}{4\pi \norm{\mathbf{g}_{u, j} - \mathbf{c}_{i}}}\bigr)^2 \nonumber \\
    &= 2(b \!+\! 1)(h_c \!-\! c_{i, 3})^b\biggl(\frac{\lambda}{4\pi}\biggr)^2 \sum_{j = 1}^{N} \frac{1}{\norm{\mathbf{g}_{u, j} \!-\! \mathbf{c}_{i}}^{b \!+\! 2}},
\end{align} and the terms $\frac{\lambda}{4 \pi}$ and $2(b+1)$ have no impact on the optimization, thus, problem \eqref{prob_MRT} can be reformulated as
\begin{subequations}\label{prob_MRT_final}
\begin{align}
\label{prob_MRT_final_a} \quad \maximize_{{t, \mathbf{g}_{u, j}, P_{u, i} v_{u,i}}} \quad & t \\ 
\textrm{subject to} \quad & \label{prob_MRT_final_b}  t   \leq \sum_{j = 1}^N \frac{e_{i}^{b} v_{u,i}P_{u, i}}{\eta_{i} \norm{\mathbf{g}_{u, j} - \mathbf{c}_{i}}^{b+2}}, \quad \forall u, i\\
& \label{prob_MRT_final_d}  \eqref{prob_basic_c}, \eqref{prob_basic_d}, \eqref{prob_basic_f}, \eqref{prob_basic_g}, \nonumber\eqref{prob_MRT_b}, 
\end{align}
\end{subequations}
where $e_{i} = h_c - c_{i, 3}$.

Problem \eqref{prob_MRT_final} can be decomposed into $U$ independent subproblems once the clustering assignment variables $v_{u,i}$ are fixed. Accordingly, the optimization is performed in two stages: first, the hotspots are partitioned into $U$ clusters by determining $v_{u,i}$; second, given this clustering, problem \eqref{prob_MRT_final} is solved separately for each cluster to optimize the deployment of its radio stripe elements. While independent per-cluster operation may be suboptimal compared to a fully coordinated multi-stripe design with joint precoding and power allocation, the framework offers a practical and scalable decentralized solution that remains tractable.

\subsection{Proposed Clustering Method}

We proceed by defining $\mathbf{s}_u$ as the location of the $u$-th cluster head, representing the central point of the $u$-th radio stripe network within its assigned cluster. Next, we define
\begin{equation}\label{PI_user}
\Delta_{u, i} = \eta_{i} \norm{\mathbf{s}_u - \mathbf{c}_{i}}^{b+2}/e_{i}^{b},
\end{equation}
as a sum path-loss metric between $i$-th hotspot and $u$-th cluster. This relies on the near-field loss by disregarding the phase term. Given the inverse relationship between received power and channel loss, and aiming to ensure fairness across receivers, we formulate the clustering problem as
\begin{subequations}\label{cluster}
\begin{align}
\label{cluster_a}
\minimize_{v_{u,i}, \mathbf{s}_{u}} \quad & \max_{u, i} v_{u,i}\Delta_{u,i}, \\
\textrm{subject to} \quad 
& \nonumber \eqref{prob_basic_f}, \eqref{prob_basic_g}.
\end{align}
\end{subequations}
Problem \eqref{cluster} is non-convex and combinatorial due to the presence of the binary assignment variables $v_{u,i} \in \{0,1\}$ and the coupled optimization over the cluster head locations $\mathbf{s}_u$. To address this challenge, we adopt an alternating optimization approach that decouples the joint optimization over the discrete assignment and continuous location variables. 

\begin{remark}
    Although \eqref{cluster} does not explicitly enforce load balancing, assigning many users to one cluster increases the penalized path loss, naturally discouraging overload. Adding explicit constraints would reduce flexibility and fairness, so the formulation balances fairness and tractability without them.
\end{remark}

\subsubsection{Optimal $\mathbf{s}_u$ given $v_{u,i}$}
 Here, we introduce an auxiliary variable $d_{u,i} = \norm{\mathbf{s}_u - \mathbf{c}_i}$ to represent the distance between the $u$-th cluster head and the $i$-th hotspot. This allows us to reformulate the problem in a more tractable form as:

\begin{subequations}\label{cluster_reform_loc}
\begin{align}
\label{cluster_reform_loc_a} 
\minimize_{t, \mathbf{s}_u, d_{u,i}} \quad & t \\ 
\textrm{subject to} \quad 
& t \geq  \eta_{i} v_{u,i} d_{u,i}^{b + 2}/e_i^b, \quad \forall u,i, \label{cluster_reform_loc_b} \\
& d_{u,i} \geq \norm{\mathbf{s}_u - \mathbf{c}_i}, \quad \forall u, i, \label{cluster_reform_loc_c}
\end{align}
\end{subequations}
where \eqref{cluster_reform_loc_c} ensures that the auxiliary variable $d_{u,i}$ correctly captures the Euclidean distance between the cluster head and the corresponding hotspot. Specifically, the inequality in \eqref{cluster_reform_loc_c} becomes tight, i.e., $d_{u,i} = \norm{\mathbf{s}_u - \mathbf{c}_i}$, at the optimum. The optimization problem in \eqref{cluster_reform_loc} is convex, as it involves a convex objective function and convex constraints. Specifically, since $d_{u,i}$ is positive, $d_{u,i}^{b + 2}$ is convex for all $b \ge -1$, which always holds. Furthermore, \eqref{cluster_reform_loc_c} is a second-order cone convex constraint, leading to the problem being a second-order conic programming (CP), solvable using standard convex optimization tools such as CVX \cite{cvxref}.

\subsubsection{Suboptimal $v_{u,i}$ given $\mathbf{s}_u$}

For fixed cluster head locations ${\mathbf{s}u}$, the hotspot-to-cluster assignment reduces to optimizing ${v{u,i}}$ under convex constraints. This leads to a mixed-integer convex program (MICP), where the objective and constraints are convex in $v_{u,i}$ but the binary nature of the variables renders the problem combinatorially hard. While exact methods such as branch-and-bound or cutting planes \cite{boyd2004convex,integerprogrammiongbook} can solve MICPs, their complexity grows exponentially with the problem size, making them impractical for large networks with many hotspots and clusters. In particular, as $U$ increases, the problem dimension escalates, often resulting in excessive runtimes or solver failures due to memory or convergence limits. To address this, we relax $v_{u,i} \in {0,1}$ to the continuous interval $[0,1]$, thereby converting the original problem into a convex linear program (LP), which can be expressed as
\begin{subequations}\label{clusterbinary}
\begin{align}
\label{clusterbinary_a} \quad \minimize_{t, {v}_{u,i}} \quad & t \\ 
\textrm{subject to} \quad & \label{clusterbinary_b}  t   \geq \eta_{i} v_{u, i}\norm{\mathbf{s}_u - \mathbf{c}_{i}}^{b + 2}/e_{i}^{b}, \quad \forall u,i, \\
&  \label{clusterbinary_c} 0 \leq v_{u,i} \leq 1, \\
&  \nonumber\eqref{prob_basic_g}.
\end{align}
\end{subequations}
The resulting fractional assignments can then be projected back to binary values using simple rounding rules or assignment heuristics, providing a suboptimal yet tractable solution.

Algorithm~\ref{cluster_alg} describes the proposed alternating optimization method for clustering. The process starts by initializing assignment variables and alternating between: (i) solving a CP to update cluster head locations, and (ii) solving an LP to update relaxed assignment variables. The process continues until convergence of the objective value. Finally, a projection step selects the maximum entry in each assignment column, setting it to one and the rest to zero.

\begin{algorithm}[t]
	\caption{Fairness-aware clustering via alternating optimization (FAC-AO)} \label{cluster_alg}
	\begin{algorithmic}[1]
		\State \textbf{Input:} $U$, $\eta_i$, $e_i$, $b$, $\epsilon$\ \textbf{Output:} $v_{u,i}$, $\mathbf{s}_u$
		\State \textbf{Initialize:} $v_{u, i}$, $t = 0$, $t' = \infty$
		\Repeat
                \State Solve \eqref{cluster_reform_loc} to update $\mathbf{s}_u$, $t' = t$
                \State  Solve \eqref{clusterbinary} to obtain $t$ and $v_{u,i}$ given $\{\mathbf{s}_u\}$
		\Until{$\left|{1 - t/{t'}}\right| < \epsilon$}
        \State \textbf{Projection:} For each $i$, set $v_{u^*,i} = 1$ where $u^* = \arg\max_u v_{u,i}$, and $v_{u,i} = 0$ for all $u \neq u^*$
	\end{algorithmic}
\end{algorithm}

For simplicity of notation, in the rest of this section, we omit the index $u$ and present the formulations for a single radio stripe network, with the understanding that the frameworks can be extended to all radio stripes in the system. Specifically, each radio stripe is assigned to exactly one cluster of hotspots, and all radio stripes operate independently. The same framework is applied to each cluster in parallel.

\subsection{SGP-based Radio Stripes Deployment}

Herein, we focus on solving the radio stripe deployment problem given a fixed clustering of the hotspots. Notably, problem~\eqref{prob_MRT_final} remains challenging due to the non-convex nature of constraints~\eqref{prob_MRT_final_b} and~\eqref{prob_basic_d}. Let us proceed by defining the auxiliary variables $d_{j, i} = \norm{\mathbf{g}_{j} - \mathbf{c}_{i}}$ and $\alpha_{j, n} = \norm{\mathbf{g}_{j} - \mathbf{g}_{n}}$ and reformulate the problem for a single radio stripe as
\begin{subequations}\label{prob_apen}
\begin{align}
\label{prob_apen_a} \quad \maximize_{\substack{\mathbf{g}_{j}, d_{j,i}, P_{i} \\t, \alpha_{j, n}}} \quad & t \\ 
\textrm{subject to} \quad & \label{prob_apen_b} e_{i}^{-b} P_{i}^{-1} t   \leq \sum_{j = 1}^N d_{j,i}^{-(b+2)}, \quad \forall i \\
& \label{prob_apen_d} \frac{1}{(N - 1)\kappa} \sum_{j = 1}^{N - 1} \alpha_{j, j+1} \leq 1, \\
&  \label{prob_apen_e} \kappa \alpha_{j,n}^{-1} \leq 1, \quad \forall j, n, \\
& \label{prob_apen_f} d_{j,i} \geq ||\mathbf{g}_{j} - \mathbf{c}_{i}||, \quad \forall j,i, \\
& \label{prob_apen_g} \alpha_{j,n} = ||\mathbf{g}_{j} - \mathbf{g}_{n}||, \forall j,n\ \text{with}\ {j\neq n}, \\
& \eqref{prob_MRT_b}, \nonumber
\end{align}
\end{subequations}
where minimizing ${d}_{j,i}\ \text{with}\ {d}_{j,i}\ge 0, \forall j,i,$ leads to maximizing $t$. Thus, \eqref{prob_apen_f} forces ${d}_{j,i}$ to be equal to the distance between hotspot $i$ and the $j$-th element of the radio stripe. 

The deployment problem in~\eqref{prob_apen} involves several constraints that fall outside the scope of standard geometric programming (GP). For example, constraint~\eqref{prob_apen_b} contains the term $\sum_{j = 1}^N d_{j,i}^{-(b+2)}$, which is a posynomial and thus not GP-compatible. Similarly, constraints~\eqref{prob_apen_f} and~\eqref{prob_apen_g} involve non-GP form distance expressions. In contrast, constraints~\eqref{prob_MRT_b},~\eqref{prob_apen_d}, and~\eqref{prob_apen_e} remain in standard GP form. To handle this structure, we employ SGP, a generalization of GP that supports both monomial and posynomial terms in the objective and constraints~\cite{gp_boyd}. While GP guarantees globally optimal solutions through convex transformation, SGP typically yields suboptimal solutions via iterative refinement. In GP, inequality constraints take the form $f(x) \leq s(x)$, where $s(x)$ must be a monomial and $f(x)$ a monomial or posynomial. SGP, however, allows both sides to be posynomials, and also extends to equality constraints involving non-monomial terms, making it well-suited for our problem. By squaring the both sides of \eqref{prob_apen_f} and \eqref{prob_apen_g}, we can write
\begin{subequations}
\begin{align}
    &d_{j,i}^{-2} \Big( \sum\nolimits_{r} {\mathbf{g}_{j}[r]}^2\! +\!  \sum\nolimits_{r} {\mathbf{c}_{i}[r]}^2 \Big)  
    \leq 1 \!+\! 2 d_{j,i}^{-2} \sum\nolimits_{r} \mathbf{g}_{j}[r] \, \mathbf{c}_{i}[r], \\
    & \alpha_{j,n}^{-2} \Big( \sum\nolimits_{r} {\mathbf{g}_{j}[r]}^2 \!+ \! \sum\nolimits_{r} {\mathbf{g}_{n}[r]}^2 \Big)
    = 1 \!+\! 2 \alpha_{j,n}^{-2} \sum\nolimits_{r} \mathbf{g}_{j}[r] \, \mathbf{g}_{n}[r],
\end{align}
\end{subequations}
which paves the way to transform the optimization problem into an SGP problem formulated as
\begin{subequations}\label{prob_sgp}
\begin{align}
\label{prob_sgp_a}  
\maximize_{\substack{
P_{i}, d_{j,i} \\
t, \alpha_{j, n}, \mathbf{g}_{j}}} 
\quad & t \\ 
\textrm{subject to} \quad 
& \label{prob_sgp_b} e_{i}^{-b} P_{i}^{-1} t \leq \sum_{j = 1}^N d_{j,i}^{-(b+2)}, \quad \forall i \\
& \label{prob_sgp_f} d_{j,i}^{-2} \biggl( \sum\nolimits_{r} {\mathbf{g}_{j}[r]}^2 +  \sum\nolimits_{r} {\mathbf{c}_{i}[r]}^2 \biggr) \nonumber \\ 
& \leq 1 + 2 d_{j,i}^{-2} \sum\nolimits_{r} \mathbf{g}_{j}[r] \, \mathbf{c}_{i}[r], \quad \forall j,i \\
& \label{prob_sgp_g} \alpha_{j,n}^{-2} \biggl( \sum\nolimits_{r} {\mathbf{g}_{j}[r]}^2 +  \sum\nolimits_{r} {\mathbf{g}_{n}[r]}^2 \biggr) \nonumber \\
& = 1 + 2 \alpha_{j,n}^{-2} \sum\nolimits_{r} \mathbf{g}_{j}[r] \, \mathbf{g}_{n}[r], \quad \forall j \neq n \\
& \label{prob_sgp_h} \eqref{prob_MRT_b}, \eqref{prob_apen_d}, \eqref{prob_apen_e} \nonumber
\end{align}
\end{subequations}

\begin{lemma}[\cite{gp_boyd}]
\label{lemma:monomial_approx}
An SGP can be approximated by a standard GP using local monomial approximation. Specifically, for a positive and differentiable function $f(\upsilon_1, \ldots, \upsilon_q)$, the best local monomial approximation near the point $(\Tilde{\upsilon}_{1}, \ldots, \Tilde{\upsilon}_{q})$ is given by
\begin{equation}\label{eq:approx}
    \hat{f}(\upsilon_1, \ldots, \upsilon_q) = f(\Tilde{\upsilon}_{1}, \ldots, \Tilde{\upsilon}_{q}) \prod\nolimits_{l = 1}^q (\upsilon_l/\Tilde{\upsilon}_{l})^{\beta_l},
\end{equation}
where the exponents $\beta_l$ are defined as
\begin{equation}\label{eq:approx_alpha}
    \beta_l = \frac{\Tilde{\upsilon}_{l}}{f(\Tilde{\upsilon}_{1}, \ldots, \Tilde{\upsilon}_{q})} \frac{\partial f(\upsilon_1, \ldots, \upsilon_q)}{\partial \upsilon_l} \bigg|_{(\upsilon_1, \ldots, \upsilon_q) = (\Tilde{\upsilon}_{1}, \ldots, \Tilde{\upsilon}_{q})}.
\end{equation}
\end{lemma}

By leveraging Lemma~\ref{lemma:monomial_approx}, problem \eqref{prob_sgp} can be transformed into a GP problem near the point $\{g^{(0)}_{j,1}, g^{(0)}_{j, 2}, d^{(0)}_{j,i}, \alpha^{(0)}_{j,n}\}$ formulated as 
\begin{subequations}\label{prob_gp}
\begin{align}
\label{prob_gp_a}  
\maximize_{\substack{
P_{i}, d_{j,i} \\
t, \alpha_{j, n}, \mathbf{g}_{j}}} \quad & t \\ 
\textrm{subject to} \quad 
& \label{prob_gp_b} e_{i}^{-b} P_{i}^{-1} t \leq \hat{h}_{i}, \quad \forall i  \\
& \label{prob_gp_f} d_{j,i}^{-2} \Big( \sum\nolimits_{r} {\mathbf{g}_{j}[r]}^2 + \sum\nolimits_{r} {\mathbf{c}_{i}[r]}^2 \Big) \leq \Tilde{h}_{j,i}, \quad \forall j,i  \\
& \label{prob_gp_g} h'_{j,n} = \bar{h}_{j,n}, \quad \forall j,n \text{ with } j \neq n \\
& \label{prob_gp_h} \frac{1}{\omega} d^{(0)}_{j,i} \leq d_{j,i} \leq \omega d^{(0)}_{j,i}, \quad \forall j,i \\
& \label{prob_gp_i} \frac{1}{\omega} \alpha^{(0)}_{j,n} \leq \alpha_{j,n} \leq \omega \alpha^{(0)}_{j,n}, \quad \forall j,n \text{ with } j \neq n \\
& \label{prob_gp_j} \frac{1}{\omega} \mathbf{g}_{j}^{(0)}[r] \leq \mathbf{g}_{j}[r] \leq \omega \mathbf{g}_{j}^{(0)}[r], \quad \forall j, r \\
& \label{prob_gp_k} \eqref{prob_MRT_b}, \eqref{prob_apen_d}, \eqref{prob_apen_e}, \nonumber
\end{align}
\end{subequations}
where $\omega$ determines the trust region of the approximation \cite{gp_boyd} and the explicit expressions for the approximated monomial functions are provided in Appendix~\ref{appendix:gp}. Notably, in case \eqref{prob_gp} becomes infeasible, one can solve the relaxed version of the problem in which \eqref{prob_gp_g} is replaced by 
\begin{equation}
    \bar{h}_{j, n}/\rho \leq h'_{j, n} \leq \rho \bar{h}_{j, n},\ \forall j,n|_{j\neq n}, 
\end{equation}
where $\rho$ is a number slightly bigger than 1 allowing the solver to violate the constraint to attain a feasible solution. Problem~\eqref{prob_gp} can be solved efficiently by standard convex optimization tools, e.g., CVX \cite{cvxref}.

\begin{algorithm}[t]
	\caption{SGP-based radio stripe deployment.} \label{sgp_alg}
	\begin{algorithmic}[1]
            \State \textbf{Input:} $\{\mathbf{g}^{(0)}_{j}\}_{\forall j}$, $\epsilon$\ \textbf{Output:} $\{\mathbf{g}_{j}\}_{\forall j}$, $\{P_{ i}\}_{\forall i}$
            \State \textbf{Initialize:} Calculate $\{d^{(0)}_{j,i}\}_{\forall j, i}$ and $\{\alpha^{(0)}_{j,n}\}_{\forall j, n}$ for $\{\mathbf{g}^{(0)}_{j}\}_{\forall j}$, $t = \inf$
            \Repeat
                \State $t' \leftarrow t$ 
                \State Solve \eqref{prob_sgp} to obtain $\{\mathbf{g}_{ j}\}$,  $\{d_{j,i}\}$, $\{\alpha_{j,n}\}$, and $t$
                \State $\mathbf{g}^{(0)}_{j} \leftarrow \mathbf{g}_{u, j}$, $d^{(0)}_{j,i} \leftarrow d_{j,i}$, $\alpha^{(0)}_{j,n} \leftarrow \alpha_{j,n}, \quad \forall j,i, n$
            \Until{$\left|{1 - t/{t'}}\right| \leq \epsilon$}
            \State Map $\mathbf{g}_{j}$ to the feasible set of \eqref{prob_MRT_final} using Algorithm~\ref{map_alg}.
\end{algorithmic} 
\end{algorithm}

Algorithm~\ref{sgp_alg} illustrates the proposed SGP-based optimization for radio stripe deployment. First, a set of locations is chosen for the elements, e.g., a square at the center of the area, and $\{{d}_{j,i}^{(0)}\}_{\forall j,i}$ and $\{{\alpha}_{j,i}^{(0)}\}_{\forall j,i}$ are initialized accordingly. Then, the solution and its neighborhood are iteratively updated until the change in the objective function becomes smaller than a specified threshold, $\epsilon$. Notice that the obtained solution might violate the practical inter-element distance constraints. For this, we propose a heuristic method to map the derived locations $\{\mathbf{g}_{j}\}_{\forall j}$ to a feasible solution for problem \eqref{prob_MRT_final}. Algorithm~\ref{map_alg} illustrates the proposed mapping for the derived solution of Algorithm~\ref{sgp_alg}. Specifically, if the distance between each pair of elements is less than $\kappa$, their distance will be increased to $\kappa$ along the direction (angle) defined by the line connecting the two elements. Moreover, the same procedure is done for any two consecutive elements in case their distance is not equal to $\kappa$. Meanwhile, always keeping the same angle between consecutive elements may lead to degenerate or looped configurations in some extreme cases, especially when initial spacing violations are severe. To improve convergence and promote geometric diversity in the layout, a small angular offset $\phi' \ll \pi/2$ is added between consecutive elements after a fixed number of iterations ($I_{\text{s}}$). This heuristic adjustment helps the mapping process escape local patterns and converge to a valid configuration.

\begin{algorithm}[t]
	\caption{Mapping the SGP solution to feasible set.} \label{map_alg}
	\begin{algorithmic}[1]
            \State \textbf{Input:} $\{\mathbf{g}_{j}\}, I_s, \phi'$\ \textbf{Output:} $\{\mathbf{g}_{j}\}$\ \textbf{Initialize:} $C = 0$
            \Repeat
            \State $C \leftarrow C + 1$
            \For{$j = 1, \ldots, N - 1$}
                \For{$n = j + 1, \ldots, N$}
                    \If{$||\mathbf{g}_{j} - \mathbf{g}_{k,n}|| > \kappa$}
                        \State $\mathbf{g}_{n}[1] \leftarrow \mathbf{g}_{j}[1] +  \frac{\mathbf{g}_{j}[1] - \mathbf{g}_{n}[1]}{\norm{\mathbf{g}_{j} - \mathbf{g}_{n}}}\kappa$ \vspace{1mm}
                        \State  $\mathbf{g}_{n}[2] \leftarrow \mathbf{g}_{j}[2] +  \frac{\mathbf{g}_{j}[2] - \mathbf{g}_{n}[2]}{\norm{\mathbf{g}_{j} - \mathbf{g}_{n}}}\kappa$ 
                    \EndIf
                \EndFor
                \vspace{1mm}
                \State $\hat{\phi} = \tan^{-1}\bigl(\frac{\mathbf{g}_{j}[2] - \mathbf{g}_{n}[2]}{\mathbf{g}_{j}[1] - \mathbf{g}_{n}[2]}\bigr)$ \vspace{1mm}
                \If{$C \mod{I_{s}} = 0$}
                     $\hat{\phi} \leftarrow \hat{\phi} + \phi'$
                \EndIf
                \If{$\norm{\mathbf{g}_{j} - \mathbf{g}_{u,j + 1}} \neq \kappa$}
                    \State $\mathbf{g}_{j + 1}[1] \leftarrow \mathbf{g}_{j}[1] +  \kappa \cos{\hat{\phi}}$
                    \State $\mathbf{g}_{j + 1}[2] \leftarrow \mathbf{g}_{j}[2] +  \kappa \sin{\hat{\phi}}$ 
                \EndIf
            \EndFor
            \Until{$\norm{\mathbf{g}_{j} - \mathbf{g}_{j + 1}} = \kappa, j = 1, \ldots, N - 1$ and $||\mathbf{g}_{j} - \mathbf{g}_{n}|| \geq \kappa, \forall j,n\ \text{with}\ {j\neq n}$}
\end{algorithmic} 
\end{algorithm}

\subsection{SCA-based Radio Stripe Deployment}

Herein, we provide a solution relying on SCA and alternating optimization for radio stripe deployment.

\subsubsection{Optimal $\{P_{i}\}$ for given $\{\mathbf{g}_j\}$}

Herein, the problem can be reformulated as
\begin{subequations}\label{prob_Pi}
\begin{align}
\label{prob_Pi_a}  \maximize_{t, \{P_{i}\}_{\forall i}} \quad & t \\ 
\textrm{subject to} \quad 
& \label{prob_Pi_b}  e_{i}^{-b} t \leq P_{i} \sum_{j = 1}^N \norm{\mathbf{g}_{j} - \mathbf{c}_{i}}^{-(b+2)}, \quad \forall i, \\
& \label{prob_Pi_c}  \sum_{i = 1}^{M_u} P_{i} \leq 1,
\end{align}
\end{subequations}
which is a standard LP and can be efficiently solved using convex optimization solvers.

\subsubsection{Suboptimal $\{\mathbf{g}_j\}$ with fixed $\{P_{i}\}$}

Here, we approximate the non-convex terms of the problem using their first-order Taylor expansions. Thus, the optimization problem near the point $\{d^{(0)}_{j,i}, \mathbf{g}^{(0)}_{j}\}_{\forall j, i}$ can be approximated for a single radio stripe as
\begin{subequations}\label{prob_SCA_Y}
\begin{align}
\label{prob_SCA_Y_a}  
\maximize_{\substack{t, \{\mathbf{g}_{j}\}_{\forall j} \\ \{d_{j, i}\}_{\forall j, i}}} \quad & t \\ 
\textrm{subject to} \quad 
& \label{prob_SCA_Y_b} e_{i}^{-b} P_{i}^{-1} t \leq \sum_{j = 1}^N \varrho_{j,i}, \quad \forall i, \\
& \label{prob_SCA_Y_d} \varrho_{j,n} \geq \kappa^2, \quad \forall j,n \text{ with } j \neq n, \\
& \label{prob_SCA_Y_e} \norm{\mathbf{g}_{j} - \mathbf{c}_{i}} \leq d_{j, i}, \quad \forall j, i, \\
& \label{prob_SCA_Y_g} \norm{\mathbf{g}_{j} - \mathbf{g}^{(0)}_{j}} \leq \sigma, \quad \forall j, \\
& \label{prob_SCA_Y_h} \eqref{prob_basic_c}, \nonumber
\end{align}
\end{subequations}
where $\sigma$ determines the trust region \cite{boyd2004convex} in which the approximation is valid, and
\begin{subequations}
\begin{align}
    &\varrho_{j,i} = {d^{(0)}_{j, i}}^{-(b+2)} -(b + 2){d^{(0)}_{j, i}}^{-(b+3)} (d_{j, i} - d^{(0)}_{j,i}), \\
    &\varrho_{j,n}  = {\norm{\mathbf{g}^{(0)}_{j} - \mathbf{g}^{(0)}_{n}}^2} + \nonumber \\
    & \quad \quad \quad \nabla^T_{\mathbf{g}_{j}} \norm{\mathbf{g}_{j} - \mathbf{g}_{n}}^2 \big|_{\substack{\mathbf{g}_{j} = \mathbf{g}^{(0)}_{j} \\ \mathbf{g}_{n} = \mathbf{g}^{(0)}_{n}}} (\mathbf{g}_{j} - \mathbf{g}^{(0)}_{j}) + \nonumber \\ 
    & \quad \quad \quad \nabla^T_{\mathbf{g}_{n}} \norm{\mathbf{g}_{j} - \mathbf{g}_{n}}^2 \big|_{\substack{\mathbf{g}_{j} = \mathbf{g}^{(0)}_{j} \\ \mathbf{g}_{n} = \mathbf{g}^{(0)}_{n}}} (\mathbf{g}_{n} - \mathbf{g}^{(0)}_{n}).
\end{align}
\end{subequations}

Notice that problem \eqref{prob_SCA_Y} is a convex program, which can be solved efficiently using standard convex optimization solvers. However, since it is based on a first-order approximation of the original non-convex constraints, the obtained solution may not be feasible for the original problem \eqref{prob_MRT_final}, particularly when the values lie near the boundary of the trust region. To overcome this, the solution of \eqref{prob_SCA_Y} is mapped to a feasible point of \eqref{prob_MRT_final} via Algorithm~\ref{map_alg}, which ensures constraint satisfaction. Moreover, the overall SCA-based radio stripe deployment is addressed through an alternating optimization procedure, as summarized in Algorithm~\ref{sca_alg}. In each iteration, the non-convex location optimization is tackled by successively refining the solution of its convex approximation \eqref{prob_SCA_Y} using SCA. Given the updated locations, the optimal power allocation ${P_i}$ is then determined via the convex LP in \eqref{prob_Pi}. This alternating process continues until convergence.

\begin{algorithm}[t]
	\caption{SCA-based alternating optimization for radio stripe deployment.} \label{sca_alg}
	\begin{algorithmic}[1]
            \State \textbf{Input:} $\{\mathbf{g}^{(0)}_{j}\}_{\forall j}$, $\epsilon$, $\chi$\ \textbf{Output:} $\{\mathbf{g}_{j}\}$, $\{P_{ i}\}$
            \State \textbf{Initialize:} Solve \eqref{prob_Pi} with $\mathbf{g}_{j} = \mathbf{g}^{(0)}_{j}, \forall j$ to obtain $t$ and $\{P_{i}\}_{\forall i}$, Calculate $\{d^{0}_{j,i}\}_{\forall j, i}$ for given $\{\mathbf{g}^{0}_j\}_{\forall j}$
            \Repeat
                \Repeat
                    \State Solve \eqref{prob_SCA_Y} to obtain $\{\mathbf{g}_{u, j}\}_{\forall j}$ \label{alg1:line:P3_start}, \quad $\mathbf{g}^{(0)}_{j} \leftarrow \mathbf{g}_{j}, \forall j$
                \Until{$\norm{\mathbf{g}_{j} - \mathbf{g}^{(0)}_{j}} \leq \chi, \forall j$}
                \State $t' \leftarrow t$, Solve \eqref{prob_Pi} to obtain $t$ and $\{P_{i}\}_{\forall i}$ 
            \Until{$\left|{1 - t/{t'}}\right| < \epsilon$}
\end{algorithmic} 
\end{algorithm}

\section{Low-Complexity Deployments}\label{sec:low_comp}

\begin{figure}[t]
\centering
    \centering
    \includegraphics[width=\columnwidth]{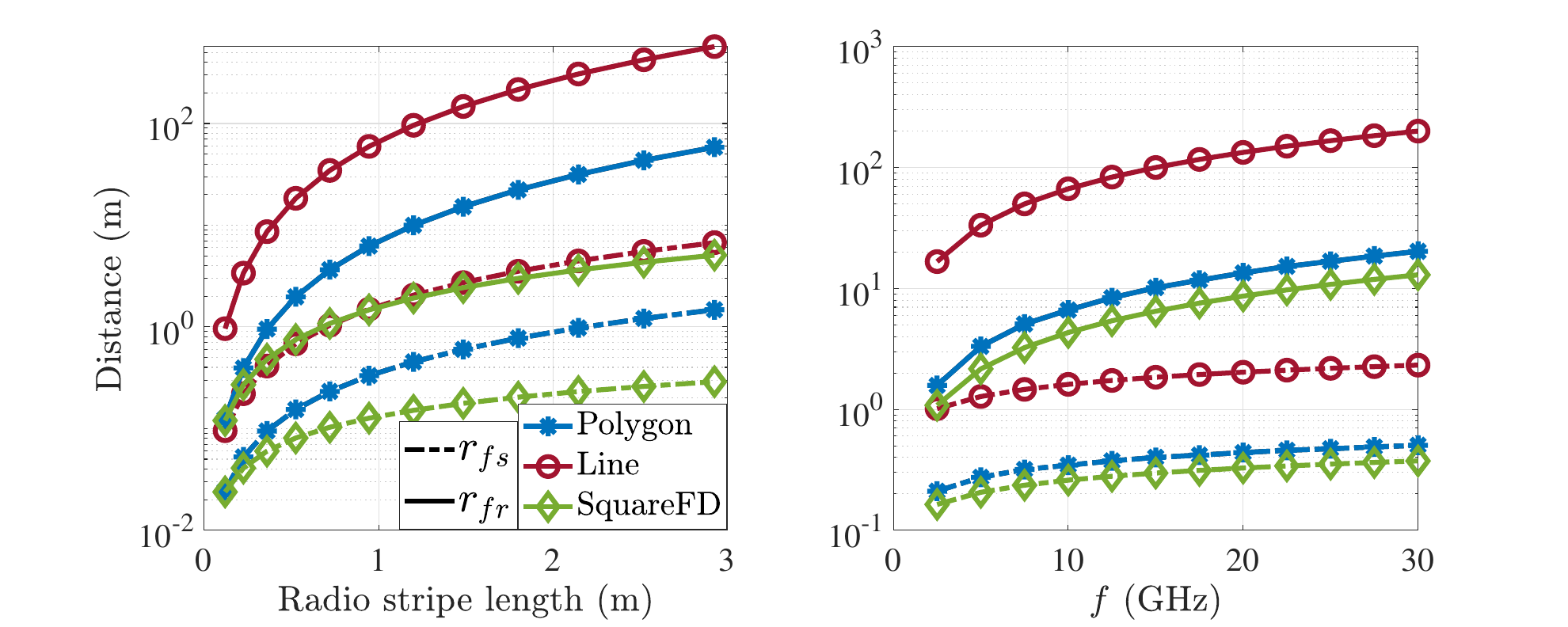}
    \caption{Fraunhofer and Fresnel distances as a function of (a) the radio stripe length with $f = 10$ GHz (left) and (b) the frequency for a 1 m radio stripe length (right). The Square-FD refers to a square uniform planar array (UPA) with its number of elements matching the nearest square number to the number of radio stripe elements for that given length/frequency.}
    \label{nearfield}
\end{figure}

One way to reduce the complexity of the deployment problem is to consider predetermined shapes, thus pre-addressing constraints \eqref{prob_basic_c} and \eqref{prob_basic_d}. Notably, using predefined shapes may be more reasonable for practical implementations as it simplifies the manufacturing and deployment of radio stripes, rather than dealing with distorted shapes. Herein, we consider two shapes for radio stripe deployment: a regular polygon and a straight line. These configurations help explore performance trade-offs between beamforming flexibility and physical layout constraints. Notably, the location of all elements in a polygon-shaped radio stripe can be mathematically modeled by specifying only the center location, while for a line-shaped deployment, it suffices to define one vertex and the orientation angle of the line. In contrast, other intermediate shapes, such as squares or rectangular loops, may require additional geometric parameters, such as side lengths or orientation, which would increase the complexity of the optimization. It is evident that for a regular polygon with the $N$ elements positioned at the vertices and denoting $r_0 = \kappa/({2\sin{\frac{\pi}{N}}})$ as the distance between the center and elements, one has $D \le 2 r_0 =\kappa/sin{\frac{\pi}{N}}$, which becomes tight as $N$ increases while converging to $\kappa N/\pi$. On the other hand, a straight line provides the largest antenna diameter, which is the length of the line, i.e., $(N - 1)\kappa$. Therefore, a straight line-shaped radio stripe makes it considerably easier to create near-field conditions. Meanwhile, one drawback of the line-shaped deployment is that some elements might have to be positioned at large distances from specific users, which can limit beam focusing precision. Moreover, line-shaped arrays have inherently limited 3D beamforming capabilities, as their narrow geometry constrains the ability to shape beams effectively in both azimuth and elevation. This limitation is less significant in polygonal deployments, which offer greater spatial coverage and angular diversity. Thus, there exists a trade-off between creating strong near-field conditions (enabled by large aperture diameters in line deployments) and achieving effective beamforming performance, depending on the deployment area and user distribution. Fig.~\ref{nearfield} visually corroborates this discussion.

\subsection{Ploygon-shaped Radio Stripe}

Assuming the polygon-shaped radio stripe, all of the elements can be positioned according to their relation with the center's location and the rotation angle. Although the location of the elements changes with rotation, the resolution of such rotation is upper bounded by $\phi = {2\pi}/{N}$, thus, it becomes negligible for a large $N$, as is typical in radio stripe systems. Hence, we discard the influence of the rotation angle and proceed by defining $\hat{\mathbf{g}}$ with $\hat{g}[3] = h_c$ as the location of the center of a regular polygon, which is placed on the ceiling. Then, we can write the location of the $j$-th element as
\begin{subequations}\label{polyeqq}
\begin{align}
    &\mathbf{g}_{j}[1] = \hat{\mathbf{g}}[1] + r_0\cos\big((j - 1)\phi\big), \\
    &\mathbf{g}_{j}[2] = \hat{\mathbf{g}}[2] + r_0\sin\big((j - 1)\phi\big).
\end{align}
\end{subequations}
 Let us proceed by defining $\hat{\mathbf{c}}_{j, i} = \mathbf{c}_{i} - [
    r_0\cos{(j - 1)\phi}, r_0\sin{(j - 1)\phi}, 0
]^T$ and $\hat{d}_{j, i} = \norm{\hat{\mathbf{g}}_k - \hat{\mathbf{c}}_{j, i}}$. Hereby, the optimization problem can be formulated as 
\begin{subequations}\label{prob_sgp_ploy}
\begin{align}
\label{prob_sgp_ploy_a} 
\maximize_{\substack{\{P_{i}, \hat{d}_{j,i}\}\\ t, \hat{\mathbf{g}}[1], \hat{\mathbf{g}}[2]}} \quad & t \\ 
\textrm{subject to} \quad 
& \label{prob_sgp_ploy_b} \eta_{i} e_{i}^{-b} P_{i}^{-1} t \leq \sum_{j = 1}^N \hat{d}_{j,i}^{-(b+2)}, \quad \forall i, \\
& \label{prob_sgp_ploy_d} \hat{d}_{j,i}^{-2} \Big( \sum\nolimits_{r} {\hat{\mathbf{g}}[r]}^{2} + \sum\nolimits_{r} {\hat{\mathbf{c}}_{j,i}[r]}^{2} \Big) \nonumber \\
& \leq 1 + 2 \hat{d}_{j,i}^{-2} \sum\nolimits_{r} \hat{\mathbf{g}}[r] \, \hat{\mathbf{c}}_{j,i}[r], \quad \forall j,i, \\
& \eqref{prob_MRT_b}, \nonumber
\end{align}
\end{subequations}
where \eqref{prob_sgp_ploy_d} comes from squaring both sides and expanding the inequality $\hat{d}_{u, j,i} \geq ||\mathbf{\hat{g}} - \hat{\mathbf{c}}_{j, i}||$. 

Notice that \eqref{prob_sgp_ploy} is an SGP, and thus can be efficiently solved by relaxing it as a GP problem, although without global optimality guarantees \cite{gp_boyd}. Herein, \eqref{prob_sgp_ploy} can be transformed into a standard GP by utilizing local monomial approximation near the point $\hat{\mathbf{g}}^{(0)}[1], \hat{\mathbf{g}}^{(0)}[2], \{\hat{d}_{j,i}^{(0)}\}$ as 
\begin{subequations}\label{prob_gp_ploy}
\begin{align}
\label{prob_gp_ploy_a}  
\maximize_{\substack{
\{P_{i}, \hat{d}_{j,i}\} \\
t, \hat{\mathbf{g}}[1], \hat{\mathbf{g}}[2]}} 
\quad & t \\ 
\textrm{subject to} \quad 
& \label{prob_gp_ploy_b} \eta_{i} e_{i}^{-b} P_{i}^{-1} t \leq \hat{h}_{i}^{(\textrm{poly})}, \quad \forall i, \\
& \label{prob_gp_ploy_d} \hat{d}_{j,i}^{-2} \Big( \sum\nolimits_{r} {\hat{\mathbf{g}}[r]}^{2} +  \sum\nolimits_{r} {\hat{\mathbf{c}}_{j,i}[r]}^{2} \Big) 
\leq \Tilde{h}_{j, i}^{(\textrm{poly})}, \ \ \forall j,i, \\
& \label{prob_gp_ploy_e} \hat{d}^{(0)}_{j,i}/\omega \leq \hat{d}_{j,i} \leq \omega \hat{d}^{(0)}_{j,i}, \ \ \forall j,i, \\
&  {\hat{\mathbf{g}}^{(0)}[r]}/\omega \leq \hat{\mathbf{g}}[r] \leq \omega \hat{\mathbf{g}}^{(0)}[r], \ \ \forall r=1,2, \\
& \eqref{prob_MRT_b}, \nonumber 
\end{align}
\end{subequations}
where $\hat{h}_{u, i}^{(\textrm{poly})}$ and $\Tilde{h}_{u, j, i}^{(\textrm{poly})}$ are calculated using \eqref{eq:hhat}-\eqref{eq:htild3} (refer to Appendix~\ref{appendix:gp}) by replacing $d_{j,i}$ and $(\mathbf{g}_{j}, \mathbf{c}_{i})$ with $\hat{d}_{j,i}$ and $(\hat{\mathbf{g}}, \hat{\mathbf{c}}_{j, i})$, respectively. Algorithm~\ref{sgp_poly_alg} illustrates the proposed optimization approach for the polygon-shaped radio stripe deployment. First, a location for the center is chosen, e.g., the center of the area, and $\{\hat{d}_{j,i}^{(0)}\}_{\forall j,i}$ is initialized accordingly. Then, the solution and its neighborhood are iteratively updated until convergence.

\begin{algorithm}[t]
	\caption{Polygon-shaped radio stripe deployment.} \label{sgp_poly_alg}
	\begin{algorithmic}[1]
            \State \textbf{Input:} $\hat{\mathbf{g}}^{(0)}[1], \hat{\mathbf{g}}^{(0)}[2]$, $\epsilon$\ \textbf{Output:} $\{\mathbf{g}_{j}\}_{\forall j}$, $\{P_{i}\}_{\forall i}$
            \State \textbf{Initialize:} Compute $\{\hat{d}^{(0)}_{j,i}\}_{\forall j,i}$ for $\hat{\mathbf{g}}^{(0)}[1], \hat{\mathbf{g}}^{(0)}[2]$, $t \gets \infty$
            \Repeat
                \State $t' \gets t$, Solve \eqref{prob_gp_ploy} to obtain $\hat{\mathbf{g}}[1], \hat{\mathbf{g}}[2]$, $\{\hat{d}_{j,i}\}_{\forall j,i}$, $t$
                \State $\hat{\mathbf{g}}^{(0)}[1] \gets \hat{\mathbf{g}}[1]$, $\hat{\mathbf{g}}^{(0)}[2] \gets \hat{\mathbf{g}}[2]$, $\hat{d}^{(0)}_{j,i} \gets \hat{d}_{j,i}, \quad \forall j,i$
            \Until{$\left|1 - {t}/{t'}\right| \leq \epsilon$}
            \State Obtain $\{\mathbf{g}_{j}\}_{\forall j}$ for $\{\hat{\mathbf{g}}[1], \hat{\mathbf{g}}[2]\}$ using \eqref{polyeqq}

\end{algorithmic} 
\end{algorithm}

\subsection{Line-shaped Radio Stripe}

Herein, we consider a straight-line-based radio stripe deployment, which is probably the most straightforward implementation. Notably, the corresponding deployment optimization problem is more complex than the polygon-shaped problem. The reason is that the horizontal angle of the line plays a crucial role and it must be optimized in addition to its center, in contrast to the regular polygon.

Let us define $\mathbf{\Tilde{g}}$ with $\Tilde{g}[3] = h_c$ as the location of the center of the line. Thus, the location of the $j$-th element is given by
\begin{subequations}\label{eq:lineloc}
\begin{align}
    \label{eq:lineloc1} \mathbf{g}_{j}[1] = \Tilde{\mathbf{g}}[1] - \big(\lfloor{N/2}\rfloor - j\big)\kappa\cos{\varphi},\\
    \label{eq:lineloc2} \mathbf{g}_{j}[2] = \Tilde{\mathbf{g}}[2] - \big(\lfloor{N/2}\rfloor - j\big)\kappa\sin{\varphi},
\end{align}
\end{subequations}
where $\varphi$ is the horizontal angle of the straight line. Then, we proceed by defining 
\begin{equation}
    \Tilde{\mathbf{c}}_{j, i} = \mathbf{c}_{i} + \biggl[
    \Big(\Big\lfloor{\frac{N}{2}\Big\rfloor} - j\Big)\kappa\cos{\varphi}, \Big(\Big\lfloor{\frac{N}{2}\Big\rfloor} - j\Big)\kappa\sin{\varphi}, 0
\biggr]^T
\end{equation}
and $\Tilde{d}_{j,i} = \norm{\Tilde{\mathbf{g}} - \Tilde{\mathbf{{c}}}_{j, i}}, \forall j,i$. Hereby, the problem becomes an SGP, which can be approximated by using \eqref{prob_gp_ploy} and replacing $\hat{q}_{j,i}$, $\hat{g}$, and $\hat{d}_{j,i}$ with $\Tilde{q}_{j,i}$, $\Tilde{g}$, and $\Tilde{d}_{j,i}$, respectively.

Moreover, the optimization procedure is similar to Algorithm~\ref{sgp_poly_alg} and consists of iteratively approximating the problem until it converges to a local optimum. Notably, the solution obtained from \eqref{prob_gp_ploy} only specifies the center of the line for a given horizontal angle; thus, the best angle must still be found since it has a huge impact on the system performance. For this, we propose the heuristic in Algorithm~\ref{alg:lineshaped}, which consists of multiple search steps to find the local optimum solution for the location of the elements. Specifically,  $\varphi$ is increased iteratively and proportionally to $\zeta \ge 1$, and the location of the center and elements for the selected angle are obtained using \eqref{prob_gp_ploy} and \eqref{eq:lineloc}, respectively. Then, the line deployment leading to the best objective function value is selected.

\section{Convergence and Complexity}\label{sec:converge}

\subsection{Convergence Analysis} Each subproblem in Algorithm~\ref{cluster_alg}, namely \eqref{cluster_reform_loc} and \eqref{clusterbinary}, is convex and solved to optimality within each iteration. The overall procedure can be viewed as a block coordinate descent method over the relaxed clustering problem (\eqref{cluster} with relaxed binary constraint), which guarantees convergence to a stationary point \cite{BCD_conv}. Although the final projection step may introduce suboptimality relative to the relaxed formulation, the resulting solution remains feasible and practically effective. Under standard assumptions, the proposed Algorithm~\ref{sgp_alg} converges to a stationary point of the surrogate convexified problem \eqref{prob_sgp} \cite{gp_boyd}. After applying the mapping step, this yields a feasible suboptimal solution for the original non-convex formulation \eqref{prob_apen}. Similarly, Algorithms~\ref{sgp_poly_alg} and \ref{alg:lineshaped} converge to a stationary point of their surrogate problem and provide a feasible suboptimal solution for \eqref{prob_sgp}. Algorithm~\ref{sca_alg} utilizes SCA to iteratively solve a non-convex but feasible and bounded problem by tackling a sequence of locally convex surrogate problems. This guarantees convergence to a stationary point of the approximated problem under standard conditions \cite{Scutari_SCA}. At each iteration, the optimal power allocation ${P_i}$ is obtained by solving a convex LP. While the resulting solution may not be globally optimal for the original non-convex formulation, the final output is a feasible and suboptimal solution to the original problem \eqref{prob_sgp}.

\subsection{Complexity Analysis}

The computational complexity of the proposed algorithms depends primarily on the number of variables and constraints in each optimization step. Let \( n \) denote the problem size, which scales with key system parameters: the number of hotspots \( K \), the number of antenna elements per stripe \( N \), and the number of clusters \( U \). Each subproblem in the proposed frameworks, whether CP, LP, or GP, can be solved using interior-point methods with a worst-case complexity of \( \mathcal{O}(n^{3.5}) \)~\cite{GP_complexity}. Therefore, while all methods share the same worst-case scaling in terms of solver operations, their practical computational load varies based on how \( n \) grows with the system size. In Algorithm~\ref{cluster_alg}, \( n \sim KU \) as the main variables include binary association indicators and cluster head positions. Algorithm~\ref{sca_alg}, which optimizes both antenna element locations and power allocation, increases the problem size to \( n \sim KN \). Algorithm~\ref{sgp_alg} introduces pairwise inter-element distance constraints in addition to per-hotspot terms, resulting in a significantly larger problem size of \( n \sim KN^2 \).  Among all, Algorithm~\ref{sgp_alg} exhibits the highest computational complexity due to its full flexibility and large number of geometric constraints. In contrast, Algorithms~\ref{sgp_poly_alg} and~\ref{alg:lineshaped} restrict the geometry to predefined layouts (polygon and line, respectively), which substantially reduces the problem size to around \( n \sim KN \). The line-shaped deployment in Algorithm~\ref{alg:lineshaped} involves an additional loop over the stripe orientation angle. Specifically, the deployment is recomputed for \(\zeta\) different angles uniformly spanning \([0, \pi]\), and for each angle, a full GP-based optimization (Algorithm~\ref{sgp_poly_alg}) is executed using the corresponding layout geometry. This adds a linear multiplicative factor to the total computational complexity. Lastly, the heuristic mapping algorithms used to enforce inter-element spacing after optimization are computationally negligible relative to the main convex subproblems and do not impact the overall complexity profile.

\begin{algorithm}[t]
	\caption{Line-shaped radio stripe deployment.} \label{alg:lineshaped}
	\begin{algorithmic}[1]
            \State \textbf{Input:} $\Tilde{\mathbf{g}}^{(0)}[1], \Tilde{\mathbf{g}}^{(0)}[2]$, $\epsilon$, $\zeta$\ \textbf{Output:} $\{\mathbf{g}^{\star}_{j}\}_{\forall j}$, $\{P_{u, i}\}_{\forall i}$
            \State \textbf{Initialize:} Compute $\{\Tilde{d}_{j, i}^{(0)}\}_{\forall i}$ for $\Tilde{\mathbf{g}}^{(0)}[1], \Tilde{\mathbf{g}}^{(0)}[2]$, $t \gets \infty$, $\Tilde{f} \gets 0$
            \For{$k = 1, \ldots, \zeta$}
                \State $\varphi \gets \frac{k\pi}{\zeta}$
                \State Call Algorithm~\ref{sgp_poly_alg} with 
     \textbf{Input:} $\Tilde{\mathbf{g}}^{(0)}[1], \Tilde{\mathbf{g}}^{(0)}[2]$, $\epsilon$  \textbf{Output:} $\{\mathbf{g}_{j}\}_{\forall j}, \{P_{u, i}\}_{\forall i}$ using \eqref{eq:lineloc} instead of \eqref{polyeqq}
                \State Compute $\min_i \norm{\mathbf{h}_{u,i}}^2$ using \eqref{eq:gain_simp}
                \If{$\min_i \norm{\mathbf{h}_{u,i}}^2 \geq \Tilde{f}$}
                    \State $\Tilde{f} \gets \min_i \norm{\mathbf{h}_{u,i}}^2$, $\mathbf{g}^{\star}_{j} \gets \mathbf{g}_{j}, \quad \forall j$
                \EndIf
            \EndFor
\end{algorithmic} 
\end{algorithm}

\section{Numerical Analysis}\label{result}

Here, we divide the analysis into two main categories: i) clustering-level analysis and ii) radio stripe-level analysis. We vary both the system frequency ($f$) and radio stripe length to assess the performance of the system under different conditions, while the optimization parameters are set to $\zeta = 50$, $\omega = 1.1$, $\epsilon = 10^{-6}$, and $\Tilde{P} = 1$ W. Furthermore, the performance indicator is the minimum received power by the users for both MRT-based and SDP-based precoders, which are obtained similarly to \cite{onellowcomp}. 

\subsection{Clustering Analysis}

We consider a $25 \times 25$ m$^2$ indoor area with $h_c = 5$ m, e.g., a shopping mall, with 25 hotspots. To reduce computational complexity in the clustering and location optimization phase, we assume that each cluster charging antenna array comprises a single element. This simplification is reasonable since the clustering process primarily determines the spatial grouping and positioning of energy transmitters relative to hotspot locations, which is largely independent of the number of antenna elements. The extension to multi-antenna stripes mainly affects the beamforming stage, which is addressed separately in the subsequent phase. Hence, the insights gained from the single-antenna setup remain applicable when each cluster is deployed using multi-antenna radio stripe networks. The total transmit power budget is $\sum_u \Tilde{P}_{u} = 1$~W, which is equally divided between the clusters, while 250 random realizations of hotspot locations are considered in the clustering analysis. Several clustering approaches have been proposed in the literature for spatial grouping of the devices in large areas. One example is the K-Means with Chebyshev centroids (K-Chebyshev) clustering algorithm \cite{onelbook}, which is effective in scenarios involving diverse distributions and distance-sensitive objectives. In this method, an initial set of cluster head locations is selected, and each hotspot is assigned to the nearest cluster head based on the Chebyshev distance metric. Then, the cluster head positions are updated to maximize the minimum distance to their associated hotspots. Although K-Chebyshev is computationally efficient, it does not explicitly account for fairness in received power, which is essential in WPT systems. Herein, we consider K-Chebyshev as a benchmark for clustering and also an initial point for Algorithm~\ref{cluster_alg}.

\begin{figure}[t]
    \centering
    \includegraphics[width=0.8\columnwidth]{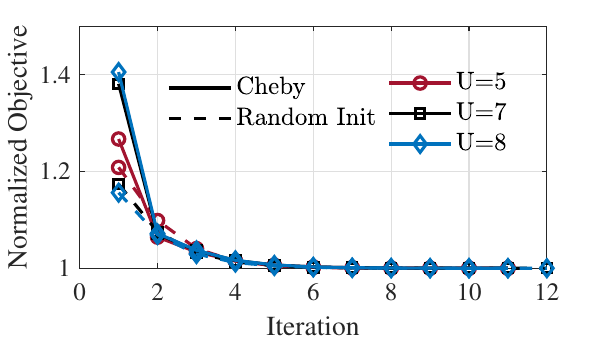}
    \caption{The evolution of the objective function in \eqref{cluster_a} with the iterations for the proposed clustering method with different initializations.}
    \label{fig:clus_conv}
\end{figure}

Fig.~\ref{fig:clus_conv} illustrates that the proposed FAC-AO algorithm converges, but its performance strongly depends on the initialization. Random initialization may cause poor user assignments and convergence to bad local minima, whereas structured initializations such as Chebyshev offer a more balanced starting point and improved performance.

\begin{figure}[t]
    \centering
    \includegraphics[width=0.8\columnwidth]{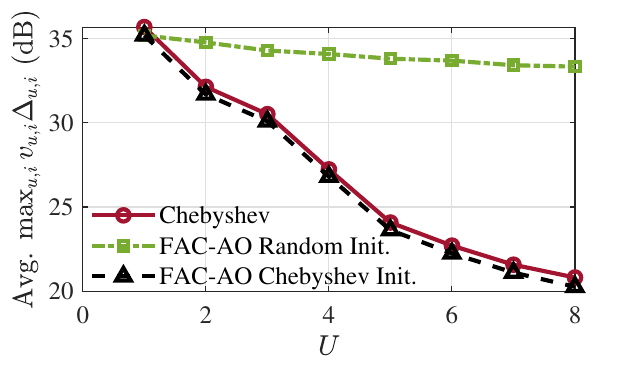}
    \caption{The average maximum loss across all hotspots as a function of $U$ assuming $\eta_i = 1, \forall i$.}
    \label{fig:clusterloss}
\end{figure}

Fig.~\ref{fig:clusterloss} shows that the average maximum loss decreases as the number of clusters increases. FAC-AO with Chebyshev initialization outperforms Chebyshev alone, while random initialization leads to poor performance, underscoring the algorithm’s sensitivity to initialization.

\begin{figure}[t]
    \centering
    \includegraphics[width=0.8\columnwidth]{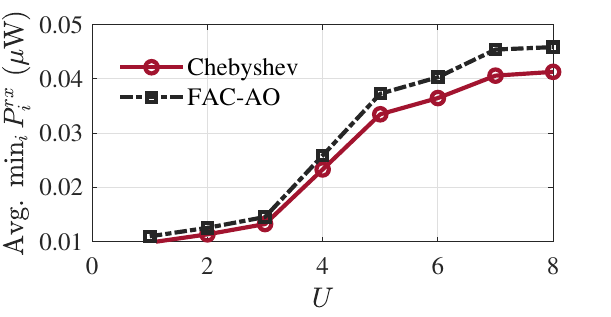}\\
    \includegraphics[width=0.8\columnwidth]{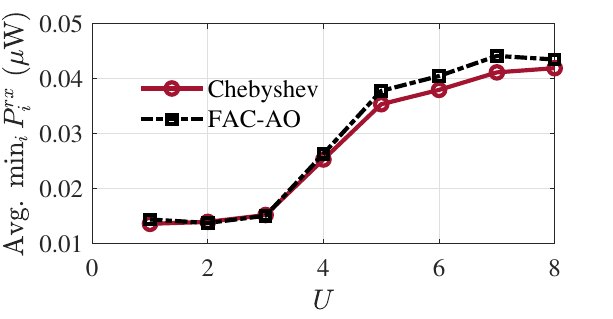}
    \caption{The average minimum received power at the center of hotspots as a function of $U$ for (a) MRT-based precoders (top) and (b) SDP-based precoders (bottom).}
    \label{fig:overcluster_power}
\end{figure}

Fig.~\ref{fig:overcluster_power} shows the average minimum received power at hotspot centers. FAC-AO outperforms Chebyshev under both MRT- and SDP-based precoders, with the performance gap growing as $U$ increases. Furthermore, increasing $U$ reduces the maximum path loss between cluster heads and hotspots, thereby improving the minimum received power.

\subsection{Radio Stripe Deployment}

To limit computational complexity, we consider a single random realization of hotspot centers, which are then clustered using FAC-AO with $U=5$. In each Monte Carlo iteration \cite{mooney1997monte}, a single user is placed randomly within a 0.5~m radius of each hotspot center, and 100~realizations of user locations are generated. For benchmarking, we compare with a UPA (Center-UPA) and a rectangle-shaped radio stripe (Center-Rectangle), both deployed at the mean Cartesian coordinates of the hotspots. The radio stripe network size is determined by its antenna length, while the UPA uses the nearest square number of elements to match the stripe’s element count.

\begin{figure}[t]
    \centering
    \includegraphics[width=0.8\columnwidth]{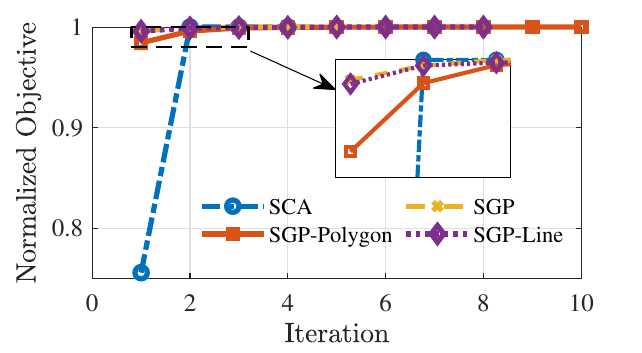}
    \caption{The evolution of the normalized objective function in \eqref{prob_apen_a} with the iterations for proposed radio stripe deployments.}
    \label{fig:RS_conv}
\end{figure}

\begin{figure}[t]
    \centering
    \includegraphics[width=0.8\columnwidth]{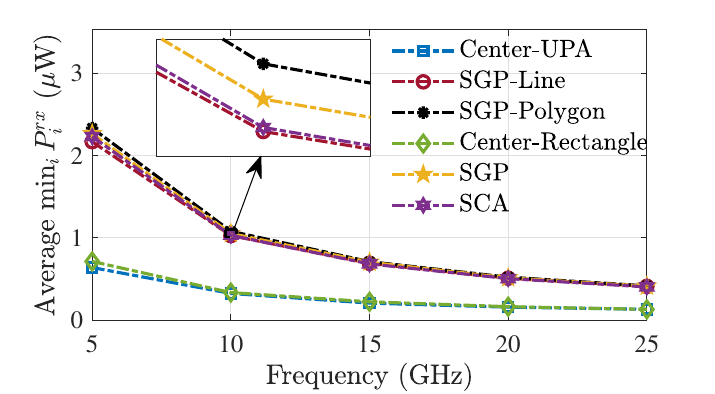}\\
    \includegraphics[width=0.8\columnwidth]{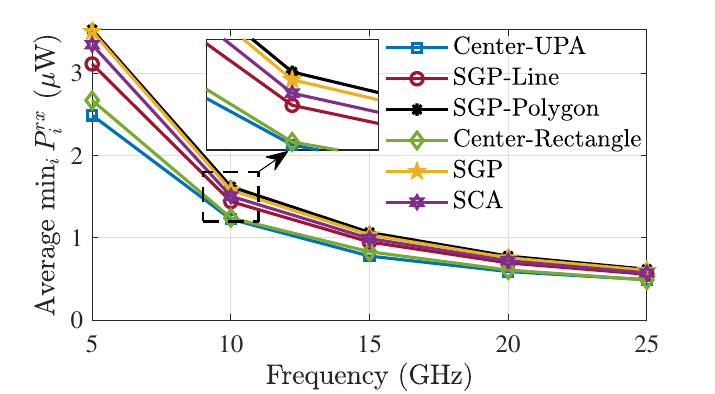}
    \caption{Average minimum power received by the users with (a) MRT-based (top) and (b) SDP-based precoders (bottom) as a function of $f$ for a 1.5~m radio stripe length.}
    \label{fig:overF}
\end{figure}

\begin{figure}[t]
    \centering
    \includegraphics[width=0.8\columnwidth]{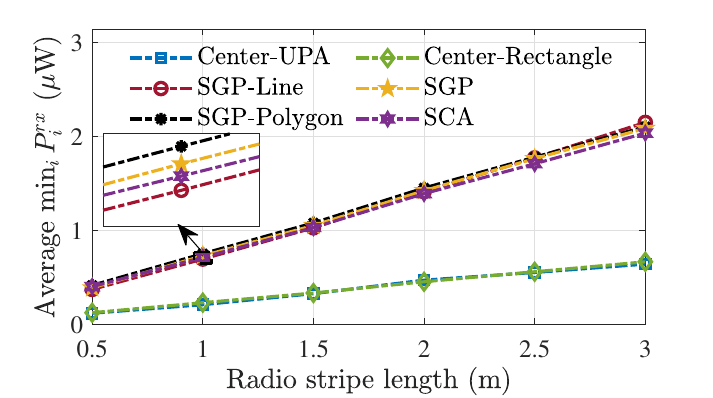}\\
    \includegraphics[width=0.8\columnwidth]{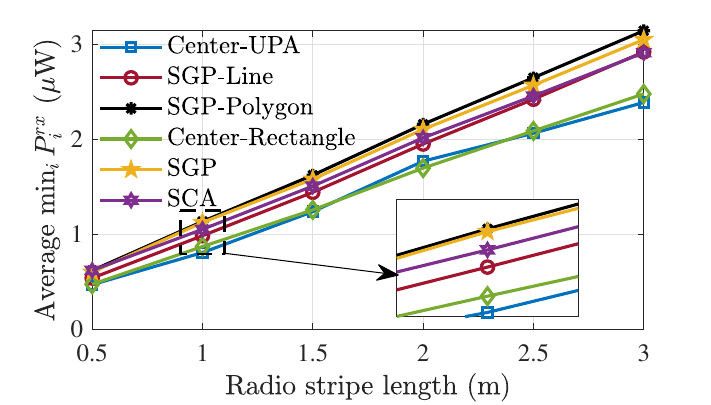}
    \caption{Average minimum power received by the users with (a) MRT-based (top) and (b) SDP-based precoders (bottom) as a function of the radio stripe length with $f = 10$~GHz.}
    \label{fig:overL}
\end{figure}

Fig.\ref{fig:RS_conv} shows that all proposed radio stripe deployment methods converge after a few iterations, confirming the convergence analysis in Section\ref{sec:converge}. Figs.~\ref{fig:overF} and \ref{fig:overL} illustrate the minimum received power versus $f$ and radio stripe length. The proposed deployments consistently outperform the benchmarks under both MRT- and SDP-based precoders, with the polygon-shaped deployment yielding the best performance by a small margin. Among the proposed methods, the line-shaped deployment performs worst, as its longer effective distances reduce the 3D beam focusing capability. While the SCA- and SGP-based approaches are not limited to predefined shapes, their reliance on approximations (and the alternating structure of SCA) restricts optimization flexibility, leading to worse performance compared to the polygon-shaped deployment, which benefits from a more direct yet geometrically constrained optimization.

As expected and seen in Fig.~\ref{fig:overL}, the performance can be improved by utilizing a larger radio stripe length since a larger number of elements can be distributed in a wider area, and more degrees of freedom are provided for beamforming. Notice that in practice, the number of RF chains also increases with the length of a fully-digital radio stripe network, thus increasing the deployment cost. On the other hand, although increasing the frequency provides more antenna elements for a given antenna length, it also increases the channel loss \cite{myEBDMA}. Moreover, since the average distance between the users and the elements does not change much over frequency, the increased losses cause a reduction in the minimum power received by the devices, as shown in Fig.~\ref{fig:overF}.

\begin{figure}[t]
    \centering
    \includegraphics[width=0.9\columnwidth]{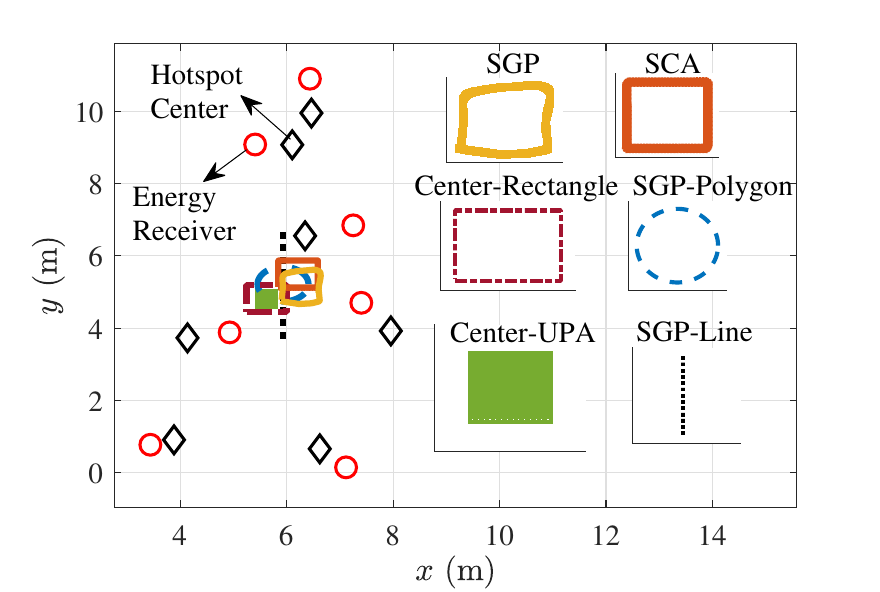}
    \caption{The 2D layout of the area illustrating the positions of both hotspots and devices, as well as the antenna deployments for $f = 10$~GHz and 1.5~m radio stripe length.}
    \label{fig:dist}
\end{figure}

Fig.~\ref{fig:dist} shows a 2D layout of one random cluster with hotspot centers, sample device positions, and antenna placements for a 3~m cable, $f=10$ GHz, and $N=200$. The highlighted radio stripe cables illustrate element distribution, with the proposed deployments positioned to cover all hotspots. The SCA-based design yields a free-form cable shape, offering flexibility and potentially higher performance but resulting in irregular layouts that may be less practical for manufacturing and installation.

\begin{figure}[t]
    \centering
    \includegraphics[width=0.8\columnwidth]{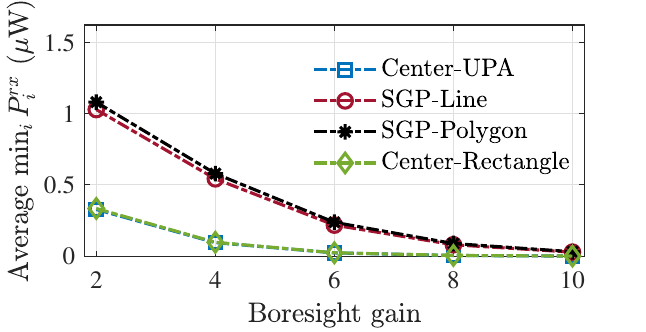}\\
    \includegraphics[width=0.8\columnwidth]{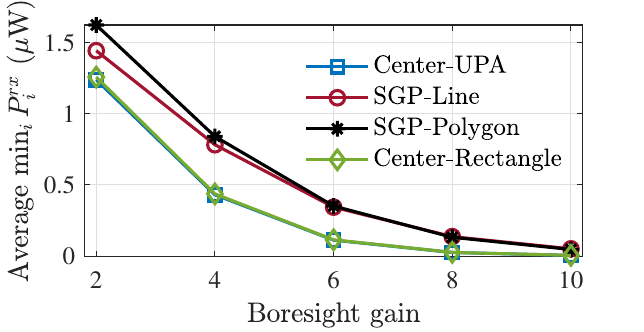}
    \caption{Average minimum power received by the users with (a) MRT-based and SDP-based precoders (top) and (b) SDP-based precoders (bottom) as a function of boresight gain of the antenna for $f = 10$~GHz and 1.5~m radio stripe length.}
    \label{fig:overb}
\end{figure}

Fig.~\ref{fig:overb} shows the impact of $b$ on system performance. Increasing $b$ generally reduces the minimum received power, as the gain becomes more focused along the boresight and coverage for off-axis users declines. Interestingly, the line-shaped deployment benefits from a larger $b$ due to its wider spatial span and angular diversity, while other deployments maintain similar orientations and cannot exploit this effect. This trend, however, is scenario-dependent: in dense single-cluster deployments, higher $b$ could enhance energy focusing. Thus, the optimal choice of $b$ depends on the deployment and user configuration and requires further investigation.

\begin{figure}[t]
    \centering
    \includegraphics[width=0.8\columnwidth]{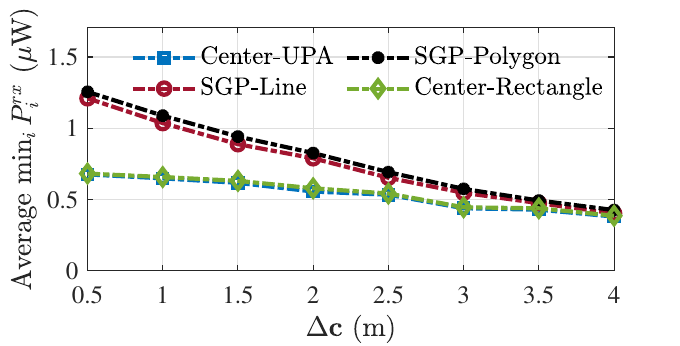}
    \caption{Average minimum received power at hotspot centers with MRT-based precoders versus $\Delta \mathbf{c}$, for $f=10$ GHz and a 1.5 m radio stripe length.}
    \label{fig:deltaC}
\end{figure}

We evaluate the robustness of different radio stripe deployments to hotspot location uncertainty by averaging the minimum received power over multiple realizations of hotspot center deviations $\Delta \mathbf{c}$ for a random cluster of 7 hotspots. Fig.~\ref{fig:deltaC} shows that performance degrades as $\Delta \mathbf{c}$ increases due to reduced alignment with transmitters, and all methods eventually converge under large deviations.

\section{Conclusions}\label{conclusion}

This paper studied radio stripe deployment for near-field WPT in indoor hotspot-centric environments, such as shopping malls. We formulated a two-stage framework that jointly optimizes clustering and antenna placement through both shape-constrained and unconstrained designs. The proposed clustering algorithm reduces worst-case path loss and outperforms the Chebyshev baseline, while all deployment strategies demonstrate convergence, with polygon-shaped layouts achieving the best overall performance. The low-complexity shape-restricted solutions strike a balance between practicality and efficiency, making them well-suited for real-world deployment.

\appendix
\section{GP-Compatible Approximations}
\label{appendix:gp}
This appendix provides the explicit monomial approximation coefficients used in the GP formulation in \eqref{prob_gp}.
\begin{align}
& \label{eq:hhat} \hat{h}_{i} = \sum_{j = 1}^N{d^{(0)}_{j,i}}^{-(b+2)} \prod_{l = 1}^N \biggl(\frac{d_{l,i}}{d^{(0)}_{l,i}} \biggr)^{\hat{\beta}_{l, i}}, \\
& \label{eq:betahhat} \hat{\beta}_{l, i} = -(b+3){d^{(0)}_{l,i}}^{-(b+2)}\Big/\Big(\sum_{j = 1}^N{d^{(0)}_{j,i}}^{-(b+2)}\Big), \\
& \label{eq:htild1} \Tilde{h}_{j, i} = \left(1 + 2{d^{(0)}_{j,i}}^{-2} \sum_{r} \mathbf{g}^{(0)}_{j}[r] \mathbf{c}_{i}[r]\right) \times \nonumber \\ 
&\hspace{20mm} \left(\frac{d_{j,i}}{d^{(0)}_{j,i}}\right)^{\Tilde{\beta}_{j, i}}  \prod_{l = 1}^{2} \left(\frac{\mathbf{g}_{j}[l]}{\mathbf{g}^{(0)}_{j}[l]}\right)^{\bar{\beta}_{j, i, l}}, \\
& \label{eq:htild2} \Tilde{\beta}_{j, i} =  \frac{-4{d^{(0)}_{j,i}}^{-2} \sum_{r} \mathbf{g}^{(0)}_{j}[r] \mathbf{c}_{i}[r]}{1 + 2{d^{(0)}_{j,i}}^{-2} \sum_{r} \mathbf{g}^{(0)}_{j}[r] \mathbf{c}_{i}[r]}, \\
& \label{eq:htild3} \bar{\beta}_{j,i, l} =  \frac{\mathbf{g}^{(0)}_{j}[l] \, {d^{(0)}_{j,i}}^{-2} \, \mathbf{c}_{i}[l]}{1 + 2{d^{(0)}_{j,i}}^{-2} \sum_{r} \mathbf{g}^{(0)}_{j}[r] \mathbf{c}_{i}[r]}, \\
&h'_{j, n} = \left({\alpha^{(0)}_{j,n}}^{-2}\left(\sum_{r} {\mathbf{g}^{(0)}_{j}[r]}^2 + \sum_{r} {\mathbf{g}^{(0)}_{n}[r]}^2\right)\right)  \times \nonumber \\ 
&\hspace{6mm} \left(\frac{\alpha_{j,n}}{\alpha^{(0)}_{j,n}} \right)^{\beta'_{j, n}} \prod_{l = 1}^2 \left(\frac{\mathbf{g}_{j}[l]}{\mathbf{g}^{(0)}_{j}[l]}\right)^{\tau_{j,n, l}} \prod_{l = 1}^2 \left(\frac{\mathbf{g}_{n}[l]}{\mathbf{g}^{(0)}_{n}[l]} \right)^{\tau'_{j,n, l}}, \\
&\beta'_{j, n} = -2, \quad \tau_{j, n, l} = \frac{2 {\mathbf{g}^{(0)}_{j}[l]}^2}{\sum_{r} {\mathbf{g}^{(0)}_{j}[r]}^2 + \sum_{r} {\mathbf{g}^{(0)}_{n}[r]}^2}, \\
&\tau'_{j,n, l} = \frac{2 {\mathbf{g}^{(0)}_{n}[l]}^2}{\sum_{r} {\mathbf{g}^{(0)}_{j}[r]}^2 + \sum_{r} {\mathbf{g}^{(0)}_{n}[r]}^2}, \\
&\bar{h}_{j, n} = \left(1 + 2 {\alpha^{(0)}_{j,n}}^{-2} \sum_{r} \mathbf{g}^{(0)}_{j}[r] \mathbf{g}^{(0)}_{n}[r]\right) \left(\frac{\alpha_{j,n}}{\alpha^{(0)}_{j,n}} \right)^{\hat{\tau}_{j,n}} \times \nonumber \\
&\hspace{24mm} \prod_{l = 1}^2 \left(\frac{\mathbf{g}_{j}[l]}{\mathbf{g}^{(0)}_{j}[l]} \right)^{\bar{\tau}_{j,l}} \prod_{l = 1}^2 \left(\frac{\mathbf{g}_{n}[l]}{\mathbf{g}^{(0)}_{n}[l]} \right)^{\Tilde{\tau}_{n,l}}, \\
& \hat{\tau}_{j,n} = \frac{-4 {\alpha^{(0)}_{j,n}}^{-2} \sum_{r} \mathbf{g}^{(0)}_{j}[r] \mathbf{g}^{(0)}_{n}[r]}{1 + 2{\alpha^{(0)}_{j,n}}^{-2} \sum_{r} \mathbf{g}^{(0)}_{j}[r] \mathbf{g}^{(0)}_{n}[r]}, \\
& \bar{\tau}_{j,n,l} = \frac{2 {\alpha^{(0)}_{j,n}}^{-2} \mathbf{g}^{(0)}_{j}[l] \, \mathbf{g}^{(0)}_{n}[l]}{1 + 2{\alpha^{(0)}_{j,n}}^{-2} \sum_{r} \mathbf{g}^{(0)}_{j}[r] \mathbf{g}^{(0)}_{n}[r]}, \\
& \Tilde{\tau}_{j, n,l} = \frac{2 {\alpha^{(0)}_{j,n}}^{-2} \mathbf{g}^{(0)}_{n}[l] \, \mathbf{g}^{(0)}_{j}[l]}{1 + 2{\alpha^{(0)}_{j,n}}^{-2} \sum_{r} \mathbf{g}^{(0)}_{j}[r] \mathbf{g}^{(0)}_{n}[r]}.
\end{align}

\ifCLASSOPTIONcaptionsoff
  \newpage
\fi

\bibliography{ref_abbv}

\begin{thebibliography}{10}

\bibitem{lópez2023highpower}
O.~Lopez~\emph{et al.}, ``{High-Power and Safe RF Wireless Charging: Cautious Deployment and Operation},'' {\em IEEE Wireless Commun.}, pp.~1--8, 2024.

\bibitem{ZEDHEXA}
O.~López~\emph{et al.}, ``Zero-energy devices for 6g: Technical enablers at a glance,'' {\em IEEE IoT Mag.}, vol.~8, no.~3, pp.~14--22, 2025.

\bibitem{intro3}
O.~L.~A. López~\emph{et al.}, ``{Massive wireless energy transfer: enabling sustainable IoT toward 6G era},'' {\em IEEE Internet Things J.}, vol.~8, no.~11, pp.~8816--8835, 2021.

\bibitem{IRS-basis}
Q.~Wu~\emph{et al.}, ``{Intelligent Reflecting Surface-Aided Wireless Communications: A Tutorial},'' {\em IEEE Trans Commun}, vol.~69, no.~5, pp.~3313--3351, 2021.

\bibitem{DAS_ref_heath}
{Heath \emph{et al.}, Robert}, ``A current perspective on distributed antenna systems for the downlink of cellular systems,'' {\em IEEE Commun. Mag.}, vol.~51, no.~4, pp.~161--167, 2013.

\bibitem{near-field}
H.~Zhang~\emph{et al.}, ``{Beam focusing for near-field multiuser MIMO communications},'' {\em IEEE Trans. Wirel. Commun.}, vol.~21, no.~9, pp.~7476--7490, 2022.

\bibitem{dist_RIS}
C.~Ma~\emph{et al.}, ``Reconfigurable distributed antennas and reflecting surface: A new architecture for wireless communications,'' {\em IEEE Trans. Commun.}, vol.~72, no.~10, pp.~6583--6598, 2024.

\bibitem{UAVzhang2018}
J.~Xu~\emph{et al.}, ``{UAV-Enabled Wireless Power Transfer: Trajectory Design and Energy Optimization},'' {\em IEEE Trans. Wirel. Commun.}, vol.~17, no.~8, pp.~5092--5106, 2018.

\bibitem{uav_survey}
X.~Gou, Z.~Sun, and K.~Huang, ``{UAV-Aided Dual-User Wireless Power Transfer: 3D Trajectory Design and Energy Optimization},'' {\em Sensors}, vol.~23, no.~6, 2023.

\bibitem{intro_radiostripe}
G.~Interdonato~\emph{et al.}, ``{Ubiquitous cell-free Massive MIMO communications},'' {\em EURASIP J Wirel Commun Netw}, vol.~2019, p.~197, Aug 2019.

\bibitem{distributedWPT1}
C.~Zhang and G.~Zhao, ``{On the Deployment of Distributed Antennas of Power Beacon in Wireless Power Transfer},'' {\em IEEE Access}, vol.~6, pp.~7489--7502, 2018.

\bibitem{Osmel_Deployment}
O.~M. Rosabal~\emph{et al.}, ``{On the Optimal Deployment of Power Beacons for Massive Wireless Energy Transfer},'' {\em IEEE Internet Things J.}, vol.~8, no.~13, pp.~10531--10542, 2021.

\bibitem{deploy_WPCN}
K.~Liang, L.~Zhao, G.~Zheng, and H.-H. Chen, ``{Non-Uniform Deployment of Power Beacons in Wireless Powered Communication Networks},'' {\em IEEE Trans. Wirel. Commun.}, vol.~18, no.~3, pp.~1887--1899, 2019.

\bibitem{WPT_dist_survey}
S.~Shen~\emph{et al.}, ``{Wireless Power Transfer With Distributed Antennas: System Design, Prototype, and Experiments},'' {\em IEEE Trans. Ind. Electron.}, vol.~68, no.~11, pp.~10868--10878, 2021.

\bibitem{twoantennaWPT}
K.~M. Mayer, L.~Cottatellucci, and R.~Schober, ``{Optimal Antenna Placement for Two-Antenna Near-Field Wireless Power Transfer},'' in {\em IEEE ICC}, pp.~2135--2140, 2023.

\bibitem{WPT_PB_indoor}
K.~M. Mayer, L.~Cottatellucci, and R.~Schober, ``{Optimal Transmit Antenna Deployment and Power Allocation for Wireless Power Supply in an Indoor Space},'' {\em IEEE Open J. Commun. Soc.}, vol.~5, pp.~3624--3640, 2024.

\bibitem{physicallylargeapperture}
B.~J. B.~D. \emph{et al.}, ``{Physically Large Apertures for Wireless Power Transfer: Performance and Regulatory Aspects},'' 2025.

\bibitem{OnelRadioStripes}
O.~L.~A. López~\emph{et al.}, ``{Massive MIMO with radio stripes for indoor wireless energy transfer},'' {\em IEEE Trans. Wirel. Commun.}, vol.~21, no.~9, pp.~7088--7104, 2022.

\bibitem{azarbahram2023radio}
A.~Azarbahram~\emph{et al.}, ``{On the Radio Stripe Deployment for Indoor RF Wireless Power Transfer},'' in {\em IEEE WCNC}, pp.~1--6, 2024.

\bibitem{anetnna_radiation}
S.~W. Ellingson, ``{Path Loss in Reconfigurable Intelligent Surface-Enabled Channels},'' in {\em IEEE PIMRC}, pp.~829--835, 2021.

\bibitem{onellowcomp}
O.~L.~A. López~\emph{et al.}, ``{A Low-Complexity Beamforming Design for Multiuser Wireless Energy Transfer},'' {\em IEEE Wireless Commun. Lett.}, vol.~10, no.~1, pp.~58--62, 2021.

\bibitem{cvxref}
M.~Grant and S.~Boyd, ``{CVX}: Matlab software for disciplined convex programming, version 2.1.'' \url{http://cvxr.com/cvx}, Mar. 2014.

\bibitem{boyd2004convex}
S.~P. Boyd and L.~Vandenberghe, {\em {Convex optimization}}.
\newblock Cambridge university press, 2004.

\bibitem{integerprogrammiongbook}
D.~Bertsimas and R.~Weismantel, {\em Optimization over integers}.
\newblock Dynamic Ideas, 2005.

\bibitem{gp_boyd}
S.~Boyd~\emph{et al.}, ``{A tutorial on geometric programming},'' {\em Optimization and Engineering}, vol.~8, pp.~67--127, Mar 2007.

\bibitem{BCD_conv}
A.~Beck and L.~Tetruashvili, ``{On the Convergence of Block Coordinate Descent Type Methods},'' {\em SIAM Journal on Optimization}, vol.~23, no.~4, pp.~2037--2060, 2013.

\bibitem{Scutari_SCA}
G.~Scutari, F.~Facchinei, and L.~Lampariello, ``{Parallel and Distributed Methods for Constrained Nonconvex Optimization—Part I: Theory},'' {\em IEEE Trans. Signal Process.}, vol.~65, no.~8, pp.~1929--1944, 2017.

\bibitem{GP_complexity}
K.~O. Kortanek, X.~Xu, and Y.~Ye, ``{An infeasible interior-point algorithm for solving primal and dual geometric programs},'' {\em Mathematical Programming}, vol.~76, pp.~155--181, Jan 1997.

\bibitem{onelbook}
H.~Alves and O.~A. Lopez, ``{Wireless RF Energy Transfer in the Massive IoT Era: Towards Sustainable Zero-energy Networks},'' 2021.

\bibitem{mooney1997monte}
C.~Z. Mooney, {\em {Monte carlo simulation}}.
\newblock No.~116, Sage, 1997.

\bibitem{myEBDMA}
A.~Azarbahram~\emph{et al.}, ``{Energy Beamforming for RF Wireless Power Transfer With Dynamic Metasurface Antennas},'' {\em IEEE Wireless Communications Letters}, vol.~13, no.~3, pp.~781--785, 2024.

\end{thebibliography}
\bibliographystyle{ieeetr}

%






\end{document}